\documentclass[aps,prd,twocolumn,nofootinbib,superscriptaddress]{revtex4-1}
\usepackage{amstext}
\usepackage{amssymb}
\usepackage{graphicx}
\usepackage{subfigure}
\usepackage{color}
\usepackage{listings}
\usepackage{natbib}
\usepackage{url}
\usepackage[colorlinks=true]{hyperref}
\usepackage{acronym}

\def\beq{\begin{equation}}\def\eeq{\end{equation}}
\def\bea{\begin{eqnarray}}\def\eea{\end{eqnarray}}

\newfont{\cursive}{pzcmi at 9pt}
\def\msun{M_{\odot}}
\def\mbh{m_{\rm BH}}
\def\mns{m_{\rm NS}}
\def\chibh{\chi_{\rm BH}}
\def\chins{\chi_{\rm NS}}

\def\newsnr{\hat{\rho}}

\begin{document}

\title{Implementing a search for aligned-spin neutron star - black hole systems with advanced ground based gravitational wave detectors}

\author{Tito \surname{Dal Canton}}
\affiliation{Max-Planck-Institut f\"ur Gravitationsphysik, Callinstrasse 38, D-30167 Hannover, Germany}
\author{Alexander H. Nitz}
\affiliation{Department of Physics, Syracuse University, Syracuse, NY 13244, USA}
\author{Andrew P. Lundgren}
\affiliation{Max-Planck-Institut f\"ur Gravitationsphysik, Callinstrasse 38, D-30167 Hannover, Germany}
\author{Alex B. Nielsen}
\affiliation{Max-Planck-Institut f\"ur Gravitationsphysik, Callinstrasse 38, D-30167 Hannover, Germany}
\author{Duncan A. Brown}
\affiliation{Department of Physics, Syracuse University, Syracuse, NY 13244, USA}
\author{Thomas Dent}
\affiliation{Max-Planck-Institut f\"ur Gravitationsphysik, Callinstrasse 38, D-30167 Hannover, Germany}
\author{Ian W. Harry}
\affiliation{Department of Physics, Syracuse University, Syracuse, NY 13244, USA}
\author{Badri Krishnan}
\affiliation{Max-Planck-Institut f\"ur Gravitationsphysik, Callinstrasse 38, D-30167 Hannover, Germany}
\author{Andrew J. Miller}
\affiliation{Abilene Christian University, Abilene, TX 79699, USA}
\affiliation{Max-Planck-Institut f\"ur Gravitationsphysik, Callinstrasse 38, D-30167 Hannover, Germany}
\author{Karl Wette}
\affiliation{Max-Planck-Institut f\"ur Gravitationsphysik, Callinstrasse 38, D-30167 Hannover, Germany}
\author{Karsten Wiesner}
\affiliation{Max-Planck-Institut f\"ur Gravitationsphysik, Callinstrasse 38, D-30167 Hannover, Germany}
\author{Joshua L. Willis}
\affiliation{Abilene Christian University, Abilene, TX 79699, USA}
\affiliation{Max-Planck-Institut f\"ur Gravitationsphysik, Callinstrasse 38, D-30167 Hannover, Germany}

\date{\today}

\begin{abstract}
  We study the effect of spins on searches for gravitational waves
  from compact binary coalescences in realistic simulated early 
  advanced LIGO data.  We construct a detection pipeline including 
  matched filtering, signal-based vetoes, a coincidence test between 
  different detectors, and an estimate of the rate of background 
  events.  We restrict attention to neutron star--black hole (NS-BH) 
  binary systems, and we compare a search using non-spinning templates 
  to one using templates that include spins aligned with the orbital 
  angular momentum.  To run the searches we implement the binary 
  inspiral matched-filter computation in PyCBC, a new software toolkit 
  for gravitational-wave data analysis.  We find that the inclusion of 
  aligned-spin effects significantly increases the astrophysical reach 
  of the search.  Considering astrophysical NS-BH systems with 
  non-precessing black hole spins, for dimensionless spin components 
  along the orbital angular momentum uniformly distributed in $(-1,
  1)$, the sensitive volume of the search with aligned-spin templates 
  is increased by $\sim 50\%$ compared to the non-spinning search; for 
  signals with aligned spins uniformly distributed in the range
  $(0.7,1)$, the increase in sensitive volume is a factor of $\sim 10$.
\end{abstract}

\maketitle

\acrodef{GW}{gravitational wave}
\acrodef{CBC}{compact binary coalescence}
\acrodef{BH}{black hole}
\acrodef{NS}{neutron star}
\acrodef{CPU}{central processing unit}
\acrodef{GPU}{graphics processing unit}
\acrodef{FFT}{fast Fourier transform}
\acrodef{PSD}{power spectral density}
\acrodef{ISCO}{innermost stable circular orbit}
\acrodef{MECO}{minimum-energy circular orbit}
\acrodef{ROC}{receiver operating characteristic}
\acrodef{SNR}{signal-to-noise ratio}

\section{Introduction}
\label{sec:intro}

We present here the first realistic \ac{GW} search
pipeline for coalescing compact binaries containing a \ac{NS}
and a \ac{BH} with spin aligned with the orbital angular momentum.
Our pipeline includes a physical template bank, signal-based vetoes and
coincidence between multiple detectors. We show that a
simple extension of traditional search methods to include the effects
of aligned spin can lead to an appreciable improvement in detection
efficiency, even during the early observational runs of advanced
\ac{GW} detectors, before they reach full sensitivity.
See \cite{Harry:2010zz,Accadia:2012zzb} for descriptions
of the advanced LIGO and Virgo detectors. In addition, the KAGRA detector
is currently under construction in Japan \cite{Somiya:2011np} and an
advanced detector has also been proposed in India \cite{LIGOIndia}.

In this paper we shall focus on neutron star--black hole (NS-BH)
binary systems which are promising sources for the advanced detectors
and pose a computational challenge. Based on our current understanding
of the population and evolution of binary systems, it is expected that
the coalescence rate for NS-BH systems within the sensitive volume of
the advanced detectors is in the range 0.2-300/year
\cite{Abadie:2010cf}. To achieve this detection rate, we must be able 
to distinguish signals from noise at a matched filter \ac{SNR} of $8$ 
or above in the LIGO detectors. Thus we require accurate models of the 
signal waveforms for matched filtering, as well as effective methods to 
exclude false alarms due to non-Gaussian artifacts in the data.

With one exception \cite{Abbott:2007ai}, previous searches of initial LIGO data
did not incorporate the effect of the compact objects' angular momentum (spin)
in the waveforms used for filtering the data\footnote{However, some other LIGO
searches (see e.g.~\cite{oai:arXiv.org:1209.6533}) have quantified how well
they could detect spinning systems, even though they conducted the primary
search using waveform models without spin.}. This was because the search
methods and detector sensitivity at the time did not warrant the inclusion of
spin effects \cite{Abbott:2007ai,VanDenBroeck:2009gd}.  In general, including
extra parameters such as spin in the search increases the size of the template
bank, making the search computationally more demanding and increasing the
false-alarm rate. An important question therefore is whether more accurate
waveform models can offset this increase in false-alarm rate. It has been
recently demonstrated that, for the case of \ac{BH}-\ac{BH} binaries, including
a single effective aligned-spin parameter in the search space does improve the
detection rate \cite{Privitera:2013xza}, but the question remains open for
NS-BH systems.  The initial LIGO detectors had relatively low sensitivity at
higher frequencies meaning that the modified phase in the expected signal due
to spin was less visible.  The situation will be different in the advanced
detector era.  The advanced LIGO detectors will be able to discern the extra
features in the waveform due to the effects of spin for a significant number of
events \cite{Brown:2012qf,Harry:2013tca}.  Furthermore, the increased
computational requirements for spinning searches can be met by improving the
analysis software used to process the data and exploiting modern computational
platforms such as \acp{GPU}.

The aim of this paper is twofold. First, we show
that it is indeed important to incorporate spin in searches for NS-BH \ac{CBC}
events, even in the early advanced detector era. Second, we describe a new
software package for \ac{CBC} searches known as PyCBC, which is designed to
meet the computational challenges of the advanced detector era.

Since the NS-BH merger rate is uncertain by about three orders of
magnitude, it is clear that much remains unknown about the population 
of compact binary systems. A measurement of this rate would constrain 
models of the formation and evolution of stellar binaries 
\cite{Mandel:2009nx,O'Shaughnessy:2005qc}.  NS-BH systems are also of 
interest astrophysically because they (along with double \ac{NS} 
systems) are expected to be progenitors of short-hard gamma-ray bursts
\cite{1989Natur.340..126E,1992ApJ...395L..83N}.  A detection of NS-BH
coalescences would allow us to explore the behavior of compact objects 
in the strong field regime and observation of the merger phase would
provide important information about the tidal disruption of \ac{NS}
and their equation of state
\cite{PhysRevD.81.064026,Pannarale:2011pk}. NS-BH systems have thus
been of significant interest for numerical relativity simulations
\cite{PhysRevD.77.084002,PhysRevD.79.044024,PhysRevD.78.104015,
  PhysRevD.74.121503,PhysRevD.74.104018}.

The spin angular momentum of binary objects affects the intrinsic evolution
of their orbits due to spin-orbit and spin-spin couplings in the post-Newtonian
orbital energy and \ac{GW} flux. If the spin of the objects in the
binary system is not aligned with
the orbital angular momentum, then the orbits will also precess
\cite{Apostolatos:1994mx}. Searches for such precessing signals are
computationally demanding as such signals are described by many 
independent 
parameters. 
Previous investigations of the effect of spin on
\ac{GW} searches largely focused on the precessing case, 
for which a number of phenomenological search templates have been
proposed \cite{Apostolatos:1995pj,Apostolatos:1996rf,Buonanno:2002fy,Grandclement:2002dv}.
However, none of these attempts were successful when applied to real data.
Here we focus instead on the simpler problem of ``aligned-spin'' 
systems, where the spin angular momenta are aligned with the orbital 
angular momentum. 

Including the effect of aligned spins still increases the size of the
template bank.  The larger number of templates
increases the number of false alarms in pure noise. False alarms from
non-Gaussian transients (glitches) triggering spinning templates had
an adverse affect on previous attempts to include spin effects in
searches \cite{VanDenBroeck:2009gd}. To counter this problem we also
include here a signal-based veto, the $\chi^2$-test
\cite{Allen:2004gu,Babak:2012zx} used in previous LIGO searches
\cite{Abbott:2009tt, oai:arXiv.org:1209.6533, Colaboration:2011np,
  Abadie:2011kd, Abadie:2010yb, Abbott:2009qj, PRD.73.102002,
  PRD.69.122001, PRD.72.082001, PRD.72.082002, 0004-637X-715-2-1453}.
This veto reduces the significance of glitch-induced triggers in the 
search, and thus greatly reduces the threshold on signal SNR that must 
be applied to achieve a desired false-alarm rate.  In order to 
simulate the behavior of real advanced LIGO data, we analyze two 
months of real data from the two 4\,km initial LIGO detectors at 
Hanford (H1) and Livingston (L1), recolored to a spectrum typical of 
the sensitivity that advanced detectors are expected to achieve in 
2016-2017 \cite{Aasi:2013wya}.

A comparison of searches with and without including spin effects
depends critically on the expected distribution of spin magnitudes and
orientations in the target astrophysical population.  The maximum
theoretical spin for an isolated Kerr \ac{BH} is given by $\chi =
1$ where $\chi$ is the dimensionless ratio $cJ/Gm^2$ between the spin
angular momentum $J$ and the mass $m$. The maximum
value of $\chi$ to which a \ac{BH} can spin up due to accretion of
matter from a thin accretion disk is thought to be very close to this
limit \cite{Thorne:1974ve}. A number of stellar mass \acp{BH} have
been discovered using X-ray techniques. Observations suggest that many
of them have quite large spins, even close to this maximum limit \cite{McClintock:2013vwa}.
This is especially true of the \acp{BH} in high-mass X-ray binaries
whose measured spins are all above $0.85$. These high-mass systems are
the most likely to form NS-BH binaries and it is likely that the 
\acp{BH} were born with these high spins since they have had insufficient
time to spin up due to accretion \cite{McClintock:2013vwa}.

Binary systems that are potential sources for advanced LIGO
are expected to have undergone a hypercritical common envelope (HCE) phase
\cite{Dominik:2012kk}. The available modeling of this phase suggests
that hypercritical accretion onto the \ac{BH} will further spin up
the \acp{BH} \cite{O'Shaughnessy:2005qc} from their spin values
before HCE. Taken in conjunction with the X-ray data, this suggests
that many of the \acp{BH} observable to the LIGO detectors will
have large spins. We test our analysis with the full range of spin values
from $-1$ to $1$ (where negative values indicate spins anti-aligned with
the orbital angular momentum) but we also display results for restricted
ranges, including a high spin range $0.7$ to $1$.

The maximum possible spin for a \ac{NS} is set by the break-up velocity,
which for expected equations of state corresponds to $\chi \sim 0.7$
\cite{Lo:2010bj}; realistic \ac{NS} values are thought to lie below this, 
braked by r-mode instabilities and perhaps by \ac{GW} emission
\cite{Wagoner:1984pv,Bildsten:1998ey,Chakrabarty:2003kt}. The maximal
spin observed for accreting millisecond pulsars corresponds to $\chi \sim 0.4$,
but the maximum value observed in a binary of two compact objects is only
$\chi \sim 0.03$ \cite{Damour:2012yf}.
In NS-BH binary systems the \ac{BH} is likely to form first due to its larger
mass and is therefore unable to contribute matter to spin up the \ac{NS}.
We therefore largely ignore the spin of the \ac{NS} in NS-BH binaries
and concentrate on a single spin, that of the \ac{BH}.

While some studies suggest that an appreciable fraction of NS-BH systems may have
significant spin misalignments \cite{Belczynski:2007xg}, others suggest a
small misalignment for most systems \cite{Kalogera:1999tq}. For small 
misalignments, the aligned-spin search would be close to optimal. 
A closed form for the waveform of a single spin precessing system
has recently been provided \cite{Lundgren:2013jla}, but a full search based
on this method has not yet been implemented and may, as an initial step,
require an efficient single-spin aligned search similar to that presented 
here. We may therefore view this investigation as the first step towards a 
fully precessing search.

A complete gravitational waveform includes the merger and post-merger ringdown
signal as well as the inspiral signal. For simplicity, we ignore the merger and
ringdown part of the waveform and ignore the possibility that the \ac{NS} may
be tidally disrupted and destroyed during the inspiral phase
\cite{Foucart:2012vn}.  A particular recent model for a complete waveform is
EOBNRv2, based on the Effective One-Body (EOB) framework calibrated by
numerical relativity waveforms \cite{Pan:2011gk}. Standard inspiral-only
waveforms were found to match EOBNRv2 waveforms for total masses below 11.4
solar masses for the advanced LIGO detectors \cite{Brown:2012nn}. Most of our
simulated signals have total masses below this limit, although it is not yet 
known what the effect of spin is on this limit. The effect of merger and 
ringdown will be studied in detail elsewhere.

The aligned-spin search pipeline employed in this paper is based on the PyCBC 
software package \cite{pycbc}. PyCBC is a newly-developed toolkit for \ac{CBC} searches
in the advanced detector era written in the Python programming language. 
It is based on modular software libraries: modules in isolation are quite simple, 
but can be put together in useful and sophisticated ways.
PyCBC allows scientists to create complicated entire end-to-end pipelines for
performing \ac{CBC} searches.  PyCBC also enables scientists to use \acp{GPU} in a
transparent manner. PyCBC builds on software from the LIGO Algorithms Library
\cite{LAL} used in previous LIGO searches.

The rest of this paper is organized as follows. Sec.~\ref{sec:pipeline}
introduces our search pipeline. Sec.~\ref{sec:pycbc} introduces the PyCBC
toolkit and the computational details of the pipeline; this section can be
read independently. The template banks are described in
Sec.~\ref{sec:templatebank} and further details of the search are in
Sec.~\ref{sec:pipelinedetails}. Sec.~\ref{sec:results} presents the main
results and Sec.~\ref{sec:conclusions} provides a summary and directions for
future work.

\section{Search method}
\label{sec:pipeline}

A fair comparison of the effects of spin in a search needs to take into
account all the details of a search of real \ac{GW} data.  Therefore we 
implement a prototype pipeline which can search for both spinning and
non-spinning systems.  This prototype pipeline is applied to a
synthetic data set obtained by recoloring initial LIGO data as
described in \cite{Aasi:2014tra}.

We first summarize the basic matched-filtering method employed,
which is described in more detail in \cite{Allen:2005fk}. 
Let $s(t)$ be the data stream from a \ac{GW} detector.  Let $n(t)$ be the
noise and $h(t)$ a \ac{GW} signal which may or may not be present in the
data stream.  Thus, $s(t)=n(t)$ in the absence of a signal, and $s(t)
= n(t) + h(t)$ otherwise.  We denote the Fourier transform of a time
series $x(t)$ as $\tilde{x}(f)$ defined as
\begin{equation}
  \tilde{x}(f) = \int_{-\infty}^\infty x(t)e^{-2\pi ift}\,dt\,.
\end{equation}
With the assumption that $n(t)$ is a stationary noise process, we define
its single-sided \ac{PSD} $S_n(f)$ as
\begin{equation}
  \langle \tilde{n}(f)\tilde{n}^\star(f^\prime)\rangle = \frac{1}{2}S_n(|f|)\delta(f-f^\prime)\,.
\end{equation}
where $\langle \cdot\rangle$ denotes the expectation value over an ensemble
of noise realizations.  While the assumption of stationarity is not a good
one for realistic data, this definition of the \ac{PSD} is still applicable
over short time scales. The non-stationarity is handled by continuously
estimating $S_n(f)$ from the data using a modification of the Welch method
\cite{PercivalWalden} as described in \cite{Allen:2005fk}.

The signal $h(t)$ as seen in the detector is a linear combination of
the two polarizations $h_+(t)$ and $h_\times(t)$:
\begin{eqnarray}
  h(t) &=& F_+(\mathbf{n},\psi;t_0)h_+(t-t_0,\phi_0) \nonumber \\
  && + F_\times(\mathbf{n},\psi;t_0)h_\times(t-t_0, \phi_0) \nonumber \\
  &=& A(t)\cos\left(\phi_0 + \phi(t-t_0)\right)\,.
\end{eqnarray}
Here the beam pattern functions $F_{+,\times}$ depend on the sky
position given by a unit-vector $\mathbf{n}$ pointing towards the
source, and on the polarization angle $\psi$ (see
e.g.~\cite{Apostolatos:1994mx}).  The beam pattern functions
$F_{+,\times}$ can be taken to be constant for the duration that the
signal is seen by the detector.  $t_0$ is a suitably defined arrival
time: in this case we will use an inspiral waveform described by the 
post-Newtonian approximation, then a convenient choice for $t_0$ is 
the termination time, such that the frequency of a signal with \ac{GW} 
phase evolution $\phi(t-t_0)$ formally becomes infinite at $t_0$. 
$\phi_0$ is the corresponding termination phase.  In the restricted 
post-Newtonian approximation, the slowly varying amplitude $A(t)$ is given by
\begin{equation}
    A(t) = -\left(\frac{G\mathcal{M}}{c^2D_{\rm eff}}\right)\left(\frac{t_0-t}{5G\mathcal{M}/c^3}\right)^{-1/4} 
\end{equation}
with $\mathcal{M} = M\eta^{3/5}$ being the chirp mass of the binary,
$M= m_1+m_2$ the total mass, $\eta = m_1m_2/M^2$ the symmetric mass ratio,
$D_{\rm eff} = D/\sqrt{F_+^2(1+\cos^2\iota)^2/4 + F_\times^2\cos^2\iota}$ the
effective distance, $\iota$ the angle between the line of sight from the
binary system to Earth and the orbital angular momentum, and $D$ the
distance to the binary system (see e.g.~\cite{Allen:2005fk}).

The termination time $t_0$ can be searched over by an inverse \ac{FFT}
and the search over $\phi_0$ can be handled by an analytic
maximization.  As shown in \cite{Allen:2005fk}, this results in having
to compute the complex statistic
\begin{equation}
  \label{eq:complexsnr}
  z(t_0) = 4\int_0^\infty \frac{\tilde{s}(f) \tilde{h}^{*}(f) }{S_{n}(f)}e^{-i2\pi ft_0}\,df\,
\end{equation}
where $\tilde{h}$ is a suitably normalized inspiral waveform template
expressed in the frequency domain. The \ac{SNR} is then defined as
$\rho = |z|/\sigma$ where
\begin{equation}
  \sigma^2 := 4\int_0^\infty \frac{|\tilde{h}(f)|^2 }{S_{n}(f)}\,df\,.
  \label{eq:sigmasq}
\end{equation}
With this normalization, in Gaussian noise in the absence of a signal we would
have $\langle\rho^2\rangle = 2$.  For practical purposes the integrations in
(\ref{eq:complexsnr}) and (\ref{eq:sigmasq}) are limited to a lower frequency
cutoff below which the detector is dominated by seismic noise and an upper
frequency cutoff beyond which the post-Newtonian waveform becomes unreliable.

\begin{figure}
    \centering
    \includegraphics[width=\columnwidth]{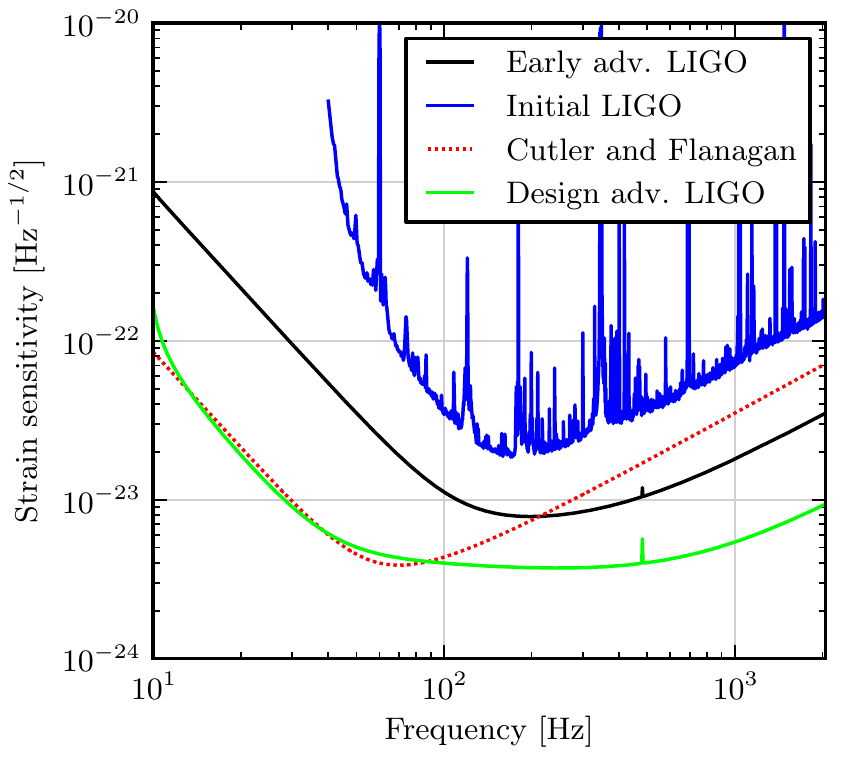}
    \caption{Sensitivity curves used in this work. The black solid
    curve corresponds to the recolored data used for testing the search.}
    \label{fig:sensitivities}
\end{figure}

This work focuses on a sensitivity curve that could reasonably represent the
early (2016) runs of advanced LIGO \cite{Aasi:2013wya} and uses a lower
frequency cutoff $f_L = 30$\,Hz. However, in Sec.~\ref{sec:templatebank} we
also investigate three different sensitivities, namely: i) the typical
sensitivity of initial LIGO during its sixth scientific run \cite{LIGO:2012aa}
with $f_L = 40$\,Hz; ii) the projected sensitivity from \cite{PhysRevD.49.2658},
used for ease of comparison with the results of \cite{Apostolatos:1996rf}, with
$f_L = 10$\,Hz; and iii) the zero-detuned, high-power design sensitivity of 
the mature advanced LIGO detectors \cite{Aasi:2013wya} with $f_L = 10$\,Hz. 
These curves are shown in
Fig.~\ref{fig:sensitivities}.

The \ac{SNR} $\rho$ works well as a detection statistic in Gaussian noise. To deal
with non-Gaussian noise of realistic detectors and veto non-Gaussian transients
of non-astrophysical origin, other statistics have been developed. A widely
used signal-based veto is the reduced $\chi^2$-statistic \cite{Allen:2004gu},
which computes the partial \acp{SNR} $\rho_{\ell}$ in $p$ non-overlapping 
frequency bands and combines them as
\begin{equation}
    \chi_{r}^{2} = \frac{p}{2p-2}\sum\limits_{\ell=1}^p\left(\rho_{\ell}- \frac{\rho}{p}\right)^2.
    \label{eq:chisquare}
\end{equation}
The bands are chosen so that a true signal with total \ac{SNR} $\rho$ would
have a partial \ac{SNR} of $\rho/p$ in each band; the union of the bands must 
cover the full frequency range used to compute $\rho$.  We note that computing
$\chi_r^2$ for each time sample requires $p$ inverse \acp{FFT} and is thus
computationally expensive.  The exact computational method of calculating 
$\rho$ and $\chi_r^2$ given a discretely sampled time series $x(t)$, the 
\texttt{FindChirp} algorithm, is described in \cite{Allen:2005fk}. We 
continue to use the same algorithm in this work.

In order to mitigate the effect of non-Gaussian transients that
plagued previous spinning studies \cite{VanDenBroeck:2009gd} we use a
modified detection statistic that extends the usual \ac{SNR} using the
$\chi^2$-veto, known as the \emph{re-weighted \ac{SNR}} statistic 
\cite{Colaboration:2011np,Babak:2012zx}:
\begin{equation}
    \newsnr = \left\{
        \begin{array}{ll}
            \rho \left[\left(1+\left(\chi_r^2\right)^3\right)/2\right]^{-1/6} & \textrm{if } \chi_r^2 > 1 \\
            \rho & \textrm{otherwise}
        \end{array}
    \right.
    \label{eq:newsnr}
\end{equation}
We threshold on both the \ac{SNR} and re-weighted \ac{SNR} when generating
candidate events, and rank them via re-weighted \ac{SNR}; this choice was 
found to be sufficient for our purposes, although it is possible that 
other choices of ranking statistic would perform even better.

\begin{figure}
  \includegraphics[width=\columnwidth]{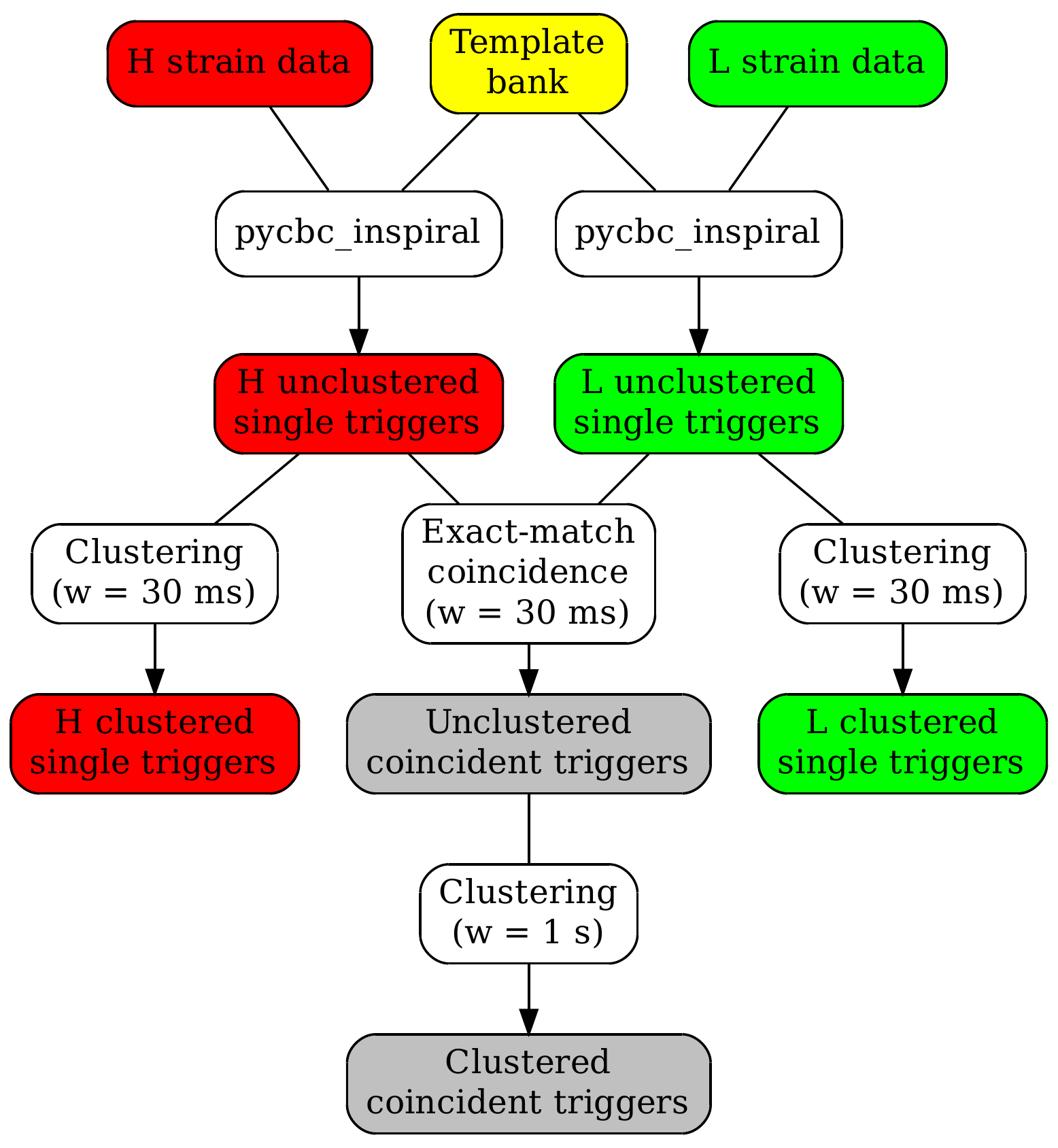}
  \caption{Flowchart of the search pipeline.  Data from the Hanford
    (H) and Livingston (L) detectors is processed by the main search
    engine \texttt{pycbc\_inspiral} which computes the \ac{SNR} and $\chi_r^2$
    time series for a common template bank which is also fixed in time.
    This results in a list of unclustered single-detector triggers which
    then pass through a
    coincidence step followed by clustering over a suitable time
    window. The single-detector triggers can additionally be clustered 
    independently in each detector.}
  \label{fig:pipeline}
\end{figure}

Our prototype search pipeline is sketched in Fig.~\ref{fig:pipeline}. 
We choose ``standard'' values for most parameters in
the pipeline (such as the number of $\chi^2$ bands $p$, the coincidence windows
etc.) that have been commonly used in other searches \cite{Babak:2012zx}.
Notable differences are the coincidence method, and the use of a template bank
which is common for all detectors and fixed for the whole data set.  A more
detailed tuning of the pipeline could improve the sensitivity further.  The
next sections introduce the PyCBC toolkit and describe the main components of
the pipeline in detail.

\section{The PyCBC Toolkit}
\label{sec:pycbc} 

Since 2004, when the first result of a \ac{CBC} search on LIGO data was
published, the bulk of the data analysis for \ac{CBC} searches has been
carried out using software from the LIGO Algorithms Library (LAL)
\cite{LAL}.  This is a set of tools and applications written in the C
programming language.  The computational landscape has diversified
significantly over the last 10 years. In particular, the use of \acp{GPU}
for general purpose computing is now more
widespread and even \ac{CPU} design is moving towards parallel architectures.
It is important that the software infrastructure for \ac{GW} searches is 
flexible enough to keep up with this diversity.

As we saw in the previous section, the computational cost for \ac{CBC} 
analyses is typically dominated by the cost of performing Fourier
transforms, primarily in computing the \ac{SNR} and the $\chi_r^2$
statistics, described earlier, for each inspiral waveform in a large
template bank.  One strategy would then be to move the \ac{FFT} 
calculations to \acp{GPU} and keep the remaining computations on the 
usual \ac{CPU} of a computer.  This would require relatively minor
modifications to the existing software in LAL and we could continue
using LAL without major modifications.  While this does speed up the
analysis time somewhat, a detailed profiling of the code reveals that
the \acp{GPU} are very under-utilized and significant time is spent in data
transfers between the \ac{CPU} and the \ac{GPU}. This suggests that further development
should allow the majority of the computation to run on the \ac{GPU}.

In this section we describe a new toolkit, PyCBC \cite{pycbc},
which builds on the tools available within LAL and makes it easier to assemble
complex end-to-end pipelines, and also enables the use of \acp{GPU} in a
transparent and user-friendly manner.  PyCBC is written in the Python
programming language \cite{python}, a convenient high-level scripting language 
with a large user community.  There are extensive collections of external
libraries in Python for a wide variety of tasks, including interfaces
to \acp{GPU} and general purpose scientific computing.  The Python modules
of PyCBC need to be able to access the existing LAL software written
in C.  This is important firstly because C can often be computationally
more efficient and secondly because LAL has an extensive collection of
\ac{GW}-specific functionality which has been well tested and widely used within
the LIGO and Virgo collaborations.

PyCBC uses the SWIG framework \cite{swig} to access LAL software for
\ac{CPU} computations.  This enables one to perform computations within PyCBC
without sacrificing computational speed. PyCBC supports \ac{GPU} computation via
either the CUDA \cite{cuda} or OpenCL \cite{opencl} architectures, using
respectively the PyCUDA \cite{pycuda} or PyOpenCL packages \cite{pyopencl}.

An example will help us illustrate how these design choices lead to
a toolkit that is flexible and maintainable, easy for users to code in,
and transparently provides the performance capability of \acp{GPU}, while also
allowing the same code to run optimally on a \ac{CPU} when that platform is
chosen instead.  A simplified script for the basic 
matched-filtering operation that performs the convolution of a template with
a data segment in PyCBC is as follows:
\begin{lstlisting}
with CUDAScheme:
  for data in segments:
    for params in bank:
      make_waveform(params, template)
      template *= data
      ifft(template, snr_time_series)
\end{lstlisting}
While a real code is somewhat more complex, particularly due to the 
thresholding, clustering, and $\chi^2$ vetoes mentioned earlier, the above
sample code shows how the design of PyCBC achieves several important goals:
\begin{enumerate}
\item Transferring data between the \ac{CPU} and \ac{GPU} is transparent to the author of
the scripts: he or she need only perform the relevant calculations inside 
the \texttt{with CUDAScheme} block (a context block in Python) and memory will
automatically be transferred as it is used in computations within the block.
In actual scripts, the context (in the example above, \texttt{CUDAScheme}) is
a variable determined at run-time, so that the same script may execute on any 
\ac{CPU}, CUDA, or OpenCL platforms.
\item We leverage Python's object oriented capabilities to ``make simple things
simple.''  In the example above, the multiplication of the template by the data
requires only the single $*\!=$ operator, though in reality it represents an
element-by-element multiplication of two frequency series, which is also
transparently sanity-checked first to ensure the two series have the same
length and frequency resolution.
\item Simplicity for the user is mirrored by comparative simplicity for PyCBC
developers, because the basic PyCBC objects (vectors, time-series, and
frequency-series) leverage the uniform interface for arithmetic and basic
mathematical operations presented by Numpy (used for \ac{CPU}), PyCUDA, and PyOpenCL.
Considering
the wide variety of basic operations, many of which can have multiple instances
depending on the precision and type (real or complex) of their inputs, this
is a huge saving in development overhead, and immediately provides a functionality
not present in LAL.
\item The inverse \ac{FFT} is transparently dispatched to the appropriate library
(CUFFT \cite{cufft} for CUDA, FFTW \cite{fftw} or MKL \cite{mkl} for \ac{CPU}) 
which again is not written by the PyCBC developers.  In the end, very little code
must be separately written for the three supported platforms; in the example
listing above, only the generation of the frequency domain waveform would be
written and maintained by the PyCBC project directly.
\end{enumerate}
As a result of this design, it is also simple to change which parts of the
computation are performed on the \ac{CPU}, and which on the \ac{GPU}. The code listing above
shows part of a script where the entire matched filter computation, and not
only the inverse \ac{FFT}, is performed on the \ac{GPU}. It therefore makes more
efficient use of the \ac{GPU} while at the same time requiring very little
additional coding.

Given the large number of templates in the spinning template bank, our
search is computationally costly; the vast majority of the cost is
represented by the matched-filtering stage, while coincidence and
clustering are comparatively trivial. The Atlas cluster \cite{atlas}
at the Albert Einstein Institute in Hannover is equipped with Nvidia
Tesla C2050 \acp{GPU} and PyCBC's flexibility allows us to accelerate
the search by running the matched-filter stage on these GPUs.  Our
implementation of the \ac{CBC} matched filtering engine uses roughly 35\%
of the \ac{GPU} time as reported by the \texttt{nvidia-smi} tool
\cite{nvidiasmi}.  A detailed profiling of the code, a performance
comparison between CPUs and GPUs, and further optimizations will be
presented elsewhere.

\section{Template banks}
\label{sec:templatebank}

To describe the template bank used in our search, we establish
some standard notation.  The inner product between two signals $h_1(t)$ and
$h_2(t)$, also known as the \emph{overlap}, is defined as
\begin{equation}
  (h_1|h_2) = 4\mathrm{Re}\int_0^\infty \frac{\tilde{h}_1(f)\tilde{h}^\star_2(f)}{S_n(f)}\,df\,.
\end{equation}
We define the normalized signal as $\hat{h}(t) := h(t)/\sqrt{(h|h)}$.
The \emph{match} between the two waveforms is defined by maximizing the
inner product between the two normalized waveforms over the time of arrival
and the phase of, say, $h_2$:
\begin{equation}
  m(h_1,h_2) = \max_{t_0,\phi_0}(\hat{h}_1|\hat{h}_2(t_0,\phi_0)) \,.
\end{equation}
Consider a template bank of $N$ waveforms $h_I$ with $I=1\ldots N$
that is meant to cover a particular parameter space of masses and spins.
The \emph{fitting factor} for any waveform $h$ in the parameter space
with the template bank is defined as:
\begin{equation}
  \label{eq:ff}
  FF = \max_I m(h_I,h)\,.
\end{equation}
In constructing a template bank, a common requirement is that any waveform
$h$ in the target parameter space must have a fitting factor
larger than 0.97~\cite{Babak:2012zx}; thus, any waveform $h$ in the 
parameter space must match some waveform in the template bank by at 
least 0.97. In the actual spinning template bank employed we find that
matches can fall as low as $0.94$. This small deterioration of the
minimal match condition occurs only in a small region of parameter
space for low values of $\eta \sim 0.05$ and will not greatly affect
signals with \ac{BH} masses below $15\msun$ and \ac{NS}
masses around $1.35\msun$. The cause of these lower match values is
discussed towards the end of this section.

In this section we compare a spinning template bank with a non-spinning
template bank. Both banks are constructed using a stochastic placement
procedure that was previously presented in \cite{Brown:2012qf}; a general
introduction to stochastic template banks can be found in
\cite{Harry:2009ea,Babak:2008rb}.  We use the stochastic bank algorithm
implemented within the PyCBC framework.  The template waveforms use the
restricted frequency-domain TaylorF2 approximant containing 3.5 pN non-spinning
phase corrections \cite{PhysRevLett.74.3515, PhysRevLett.93.091101} and 2.5 pN
spinning phase corrections \cite{Arun:2008kb, PhysRevD.49.2658,
PhysRevD.52.821, PhysRevD.74.104034}.  When calculating the matched-filter
\ac{SNR}, our template waveforms terminate at a frequency corresponding to the
\ac{ISCO} of a Schwarzschild \ac{BH} of the same total mass as the template,
i.e.~$f_{\rm ISCO} := c^3 (6\sqrt{6}\pi G M)^{-1}$. This was the standard
choice in past \ac{CBC} searches. However, in the construction of our banks,
templates are assumed to terminate at a fixed frequency of 1000 Hz, which is
close to the maximum \ac{ISCO} frequency in our parameter space. Past searches
also made a fixed-frequency assumption. Although PyCBC has the ability to
construct banks with a varying termination frequency, we do not explore the
effect of this choice in this study.

The template bank for the non-spinning search has a \ac{BH} mass $\mbh$
ranging from 3 to 15 $\msun$ and a \ac{NS} mass $\mns$ ranging from 1 $\msun$
to the equal-mass boundary $\mbh = \mns$.  We also impose the
constraint $M \le 18$ $\msun$. Both spins are
constrained to zero. This results in $\sim 28000$ templates. The bank
for the spinning search is constructed instead with $\mbh \in [3,15]$
$\msun$, $\mns \in [1,3]$ $\msun$, $\chibh \in [-1,1]$ and $\chins \in
[-0.4,0.4]$. Such settings produce $\sim 150\ 000$ templates, which
turn out to be mostly clumped around extremal values of $\chibh$.

\begin{figure}
\centering
 \includegraphics[width=\columnwidth]{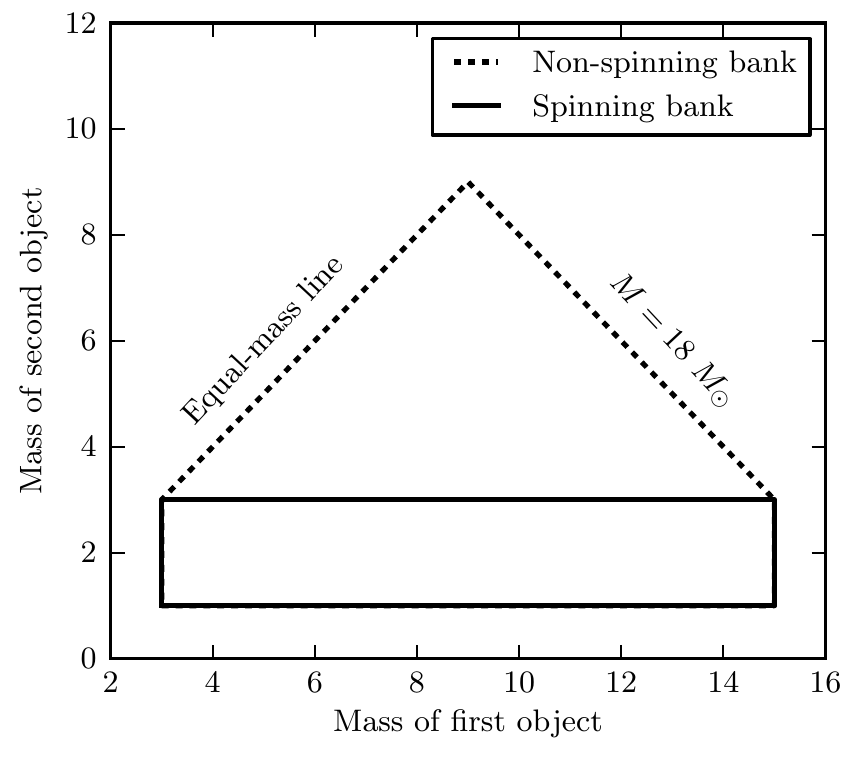}
 \caption{Mass boundaries used in constructing our non-spinning and
   spinning template banks. The non-spinning bank includes templates
   with \ac{NS} masses above the usual \ac{NS} mass range. As explained in the text,
   these templates are able to detect spinning NS-BH signals with a \ac{NS}
   mass in the usual range.}
 \label{fig:massboundaries}
\end{figure}

Fig.~\ref{fig:massboundaries} shows the mass boundary of the two
banks. As can be seen, the non-spinning bank has a larger mass range
for the \ac{NS} than the spinning bank, in particular it includes part 
of the binary \ac{BH} region. We make this choice partly because
this is how a traditional low-mass non-spinning search would be
carried out and partly to allow spinning signals to be recovered by
non-spinning templates with similar chirp mass but closer to the
equal-mass boundary, thanks to a degeneracy between spin and symmetric
mass ratio \cite{PhysRevD.87.024035}.
In other words, we explicitly favor the non-spinning search by tolerating
a bias in the recovered symmetric mass ratio. The fraction
of templates in the non-spinning bank with $\mns > 3 \msun$ is $\sim 6\%$.

\begin{figure}
\centering
 \includegraphics[width=\columnwidth]{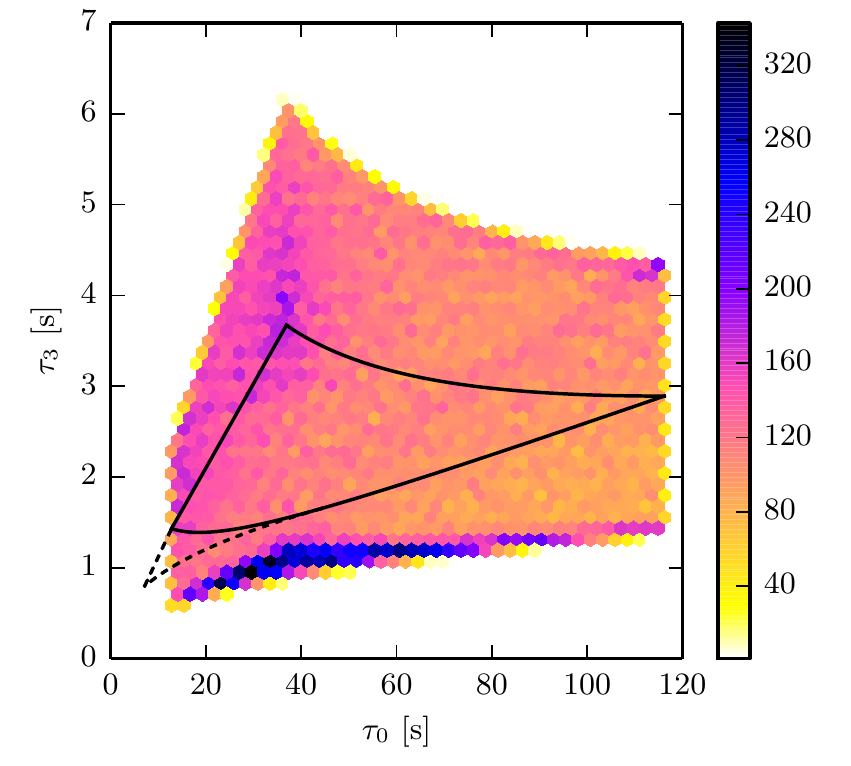}
 \caption{Template density of the spinning stochastic bank in $(\tau_0, \tau_3)$
    coordinates. The black contour delimits the non-spinning region of NS-BH
    parameter space and the dashed lines show the additional \ac{NS} mass range
    allowed by the non-spinning bank. Templates above and below the black
    contour correspond to $\chibh < 0$ and $\chibh > 0$ respectively.}
 \label{fig:templatebank}
\end{figure}

Fig.~\ref{fig:templatebank} shows the template density of the
spinning bank in the $(\tau_0,\tau_3)$ plane, where
\begin{eqnarray}
  \tau_0 &=& \frac{5}{256\pi\eta f_0}(\pi M f_0)^{-5/3} \\
  \tau_3 &=& \frac{1}{128\pi\eta f_0}(\pi Mf_0)^{-2/3} \times \nonumber \\
  && \times \left[16\pi - \frac{\chibh}{6}(19\delta^2 + 113\delta + 94)\right]
\end{eqnarray}
are the \emph{chirp times} \cite{Sengupta:2003wk} extended to include
spin-orbit effects, $\delta = (\mbh - \mns) / M$ and $f_0 = 20$
Hz is a fiducial frequency.  The non-spinning part of the NS-BH parameter
space is shown as a black contour in the figure and the region covered
by the non-spinning bank corresponds to the black contour plus the
small area delimited by the dashed contour. As can be seen, including
the effect of spin broadens the covered region significantly. Moreover,
although the density remains approximately
constant inside the black contour, it increases noticeably outside; in
particular, a large amount of templates is concentrated at small
$\tau_3$ values.  Better coordinates for representing spinning
templates in which the template density is nearly constant are given in
\cite{Brown:2012qf,Ohme:2013nsa}.

\subsection{Fitting-factor calculations}
\label{subsec:banksim}

The behavior of a template bank with respect to various signals can be studied
without the effect of detector noise by numerically evaluating the fitting
factors defined in (\ref{eq:ff}), which can be done by PyCBC.  In order to get
a first characterization of the effect of spin on a few non-spinning banks
associated with the different sensitivity curves and lower-frequency cutoffs
shown in Fig.\ref{fig:sensitivities}, we calculate the fitting factors for such
banks using simulated signals with fixed masses ($\mbh = 7.8\,\msun$ and $\mns
= 1.35\,\msun$) and a full range of physical spins for the heavier object ($-1
< \chibh < 1$) and zero spin for the lighter object ($\chins = 0$). The signals
are simulated using the standard time-domain SpinTaylorT2 approximant available
in LALSimulation \cite{LAL}. In principle one could choose other approximants
such as SpinTaylorT1 or SpinTaylorT4, which treat
the Taylor expansions of the energy and flux differently. However, we choose
SpinTaylorT2 because it is essentially the time-domain version of our
frequency-domain templates, reducing issues related to agreement between signal
and template approximants which are outside the scope of this paper. The
waveform generation starts at 20 Hz (well outside the integration range of the
matched filter) and terminates at the \ac{MECO} after which the evolution of
the orbit is no longer expected to be adiabatic (see e.g.~\cite{Pan:2003qt}).
This choice is different from the termination condition assumed in the
construction of the template banks (1000\,Hz) as well as the upper frequency
limit used in matched filtering (the template \ac{ISCO} frequency). In reality,
a physical NS-BH waveform terminates with the merger and ringdown, typically at
frequencies higher than \ac{ISCO}, so any choice of abrupt termination of the
signal is artificial. Given that we do not consider NS-BH merger and ringdown
in this study, \ac{MECO} is a good choice both for implementation reasons and
because it is also almost always greater than \ac{ISCO}. Nevertheless, as
discussed later in this section, this discrepancy can affect the fitting
factor of binaries at high mass or high positive \ac{BH} spin. The performance
of our template banks for more realistic signal models including merger and
ringdown will be assessed in a future study.

\begin{figure}
\centering
 \includegraphics[width=\columnwidth]{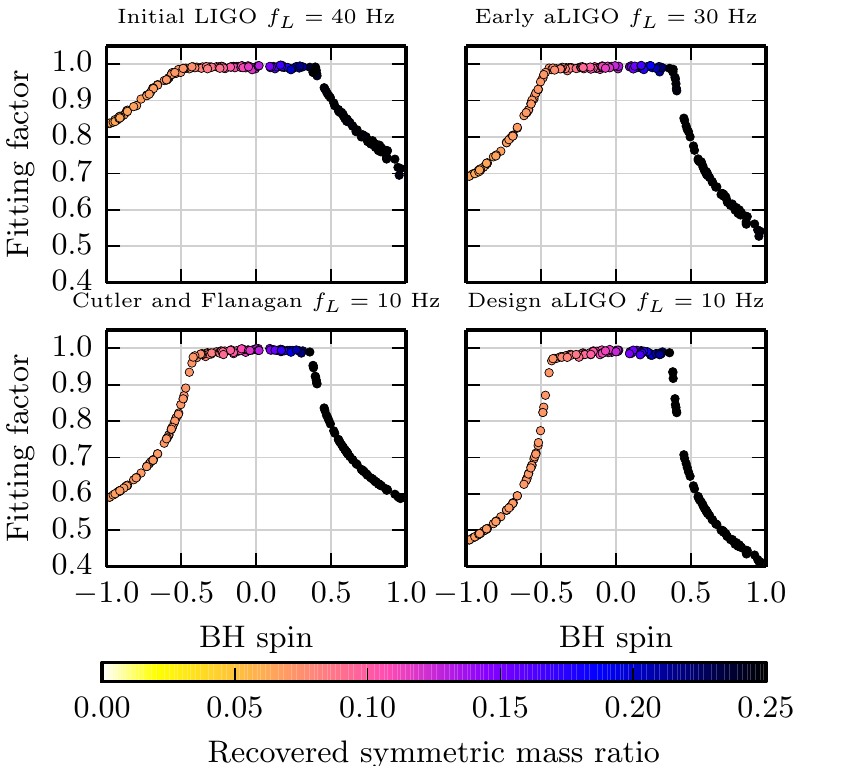}
 \caption{Fitting factor of SpinTaylorT2 NS-BH signals with fixed masses
    ($\mbh = 7.8$ $\msun$, $\mns = 1.35$ $\msun$) and different \ac{BH}
    spins and non-spinning template banks constructed for different 
    choices of sensitivity and lower cutoff frequency $f_L$
    (see Fig.\ref{fig:sensitivities}). The non-spinning bank considered 
    in the rest of this paper corresponds to the early advanced-LIGO 
    sensitivity (top right plot).}
 \label{fig:nospinmatches}
\end{figure}

The results for the different non-spinning banks are given in
Fig.~\ref{fig:nospinmatches}, showing similar behavior over different
sensitivity curves and lower frequency cutoffs.
As can be seen, in all banks there is a range of low \ac{BH} spins for which
the non-spinning bank is able to match the spinning signals fairly well, but
then a sharp fall-off in the fitting factor occurs above $|\chibh| \sim 0.4$.
The shading (color online) of the points shows the recovered value of $\eta$ in
the non-spinning bank. Although the signals are simulated with $\eta=0.126$, as
the spin is increased the recovered value of $\eta$ also increases,
compensating for the larger spin. It can be seen that the sharp fall-off in the
fitting factor for positively aligned systems is associated with the maximum
physical value of $\eta=1/4$, corresponding to equal mass templates.  Thus, if
we had injected signals with a different value of $\eta$, the fall-off in match
could happen at different values of $\chi$. For the equal mass case $\eta=1/4$,
for instance, we are already at the boundary and $\eta$ cannot increase any
further to compensate for the spin. The match then starts to decrease sharply
even for small positive spins. In the rest of the paper we will only consider
the early advanced LIGO sensitivity curve (top-right panel of
Fig.~\ref{fig:nospinmatches}) and references to ``the (non-)spinning bank''
will denote banks built for this case.

\begin{figure}
\centering
 \includegraphics[width=\columnwidth]{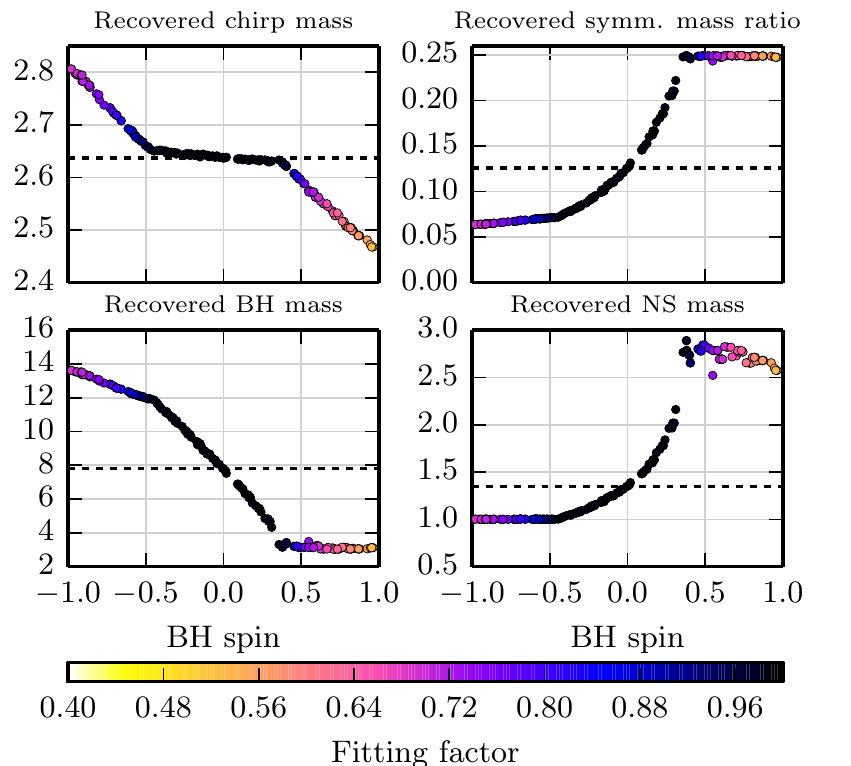}
 \caption{Mass parameters recovered in the non-spinning bank for
   SpinTaylorT2 signals with fixed masses ($\mbh = 7.8\,\msun$,
   $\mns = 1.35\,\msun$, dashed lines) and variable dimensionless
   \ac{BH} spin (x axes).}
 \label{fig:nospinmasses}
\end{figure}

The loss of match at high \ac{BH} spins can be further understood by
comparing the true and recovered values of the masses in the
non-spinning bank, as is done in Fig.~\ref{fig:nospinmasses}.
The shading now shows the match and again it is clear that the
rapid fall-off in match for the positively aligned waveforms is due to
the boundary at $\eta=1/4$. The rapid fall-off in the match for
anti-aligned waveforms is due to a different effect, namely the fact
that the minimum $\mns$ in all template banks is 1 $\msun$. Unlike the
$\eta=1/4$ case, this is not a physical boundary and one could obtain
better matches for highly spinning anti-aligned systems by lowering
the minimum \ac{NS} mass in the template bank.

\begin{figure}
\centering
 \includegraphics[width=\columnwidth]{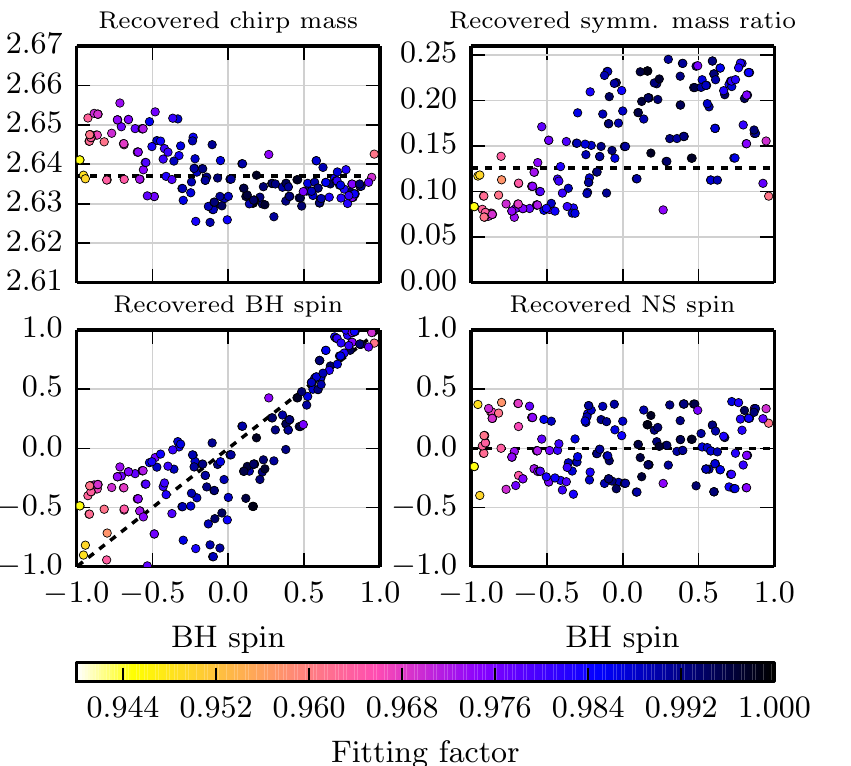}
 \caption{Mass and spin parameters recovered in the spinning bank for
   SpinTaylorT2 signals with fixed masses ($\mbh = 7.8\,\msun$,
   $\mns = 1.35\,\msun$) and variable dimensionless \ac{BH} spin
   (x axes). Dashed lines show the true parameters.}
 \label{fig:spinmasses}
\end{figure}

The recovered mass and spin parameters in the spinning bank are given in
Fig.~\ref{fig:spinmasses}.
Here, as expected the templates are all well-matched, although there
is a slight bias in the recovered masses and $\chibh$ values. The
recovered $\chins$ value is seen to be widely scattered and it is
clear that this does not have any significant impact on the match, nor
is its value well recovered by the bank.  In other aligned-spin search
investigations \cite{Privitera:2013xza} a single effective spin
parameter was used and the minimal impact of the $\chins$ value seen
here is consistent with that approach. The fact that the matches
descend below $0.95$ despite the design choice that the bank should
have a minimal match of $0.97$ was also noted in \cite{Harry:2013tca}
and was explained there by an inconsistency between the termination
condition of template and signal waveforms.

\begin{figure*}
    \centering
    \includegraphics[width=2\columnwidth]{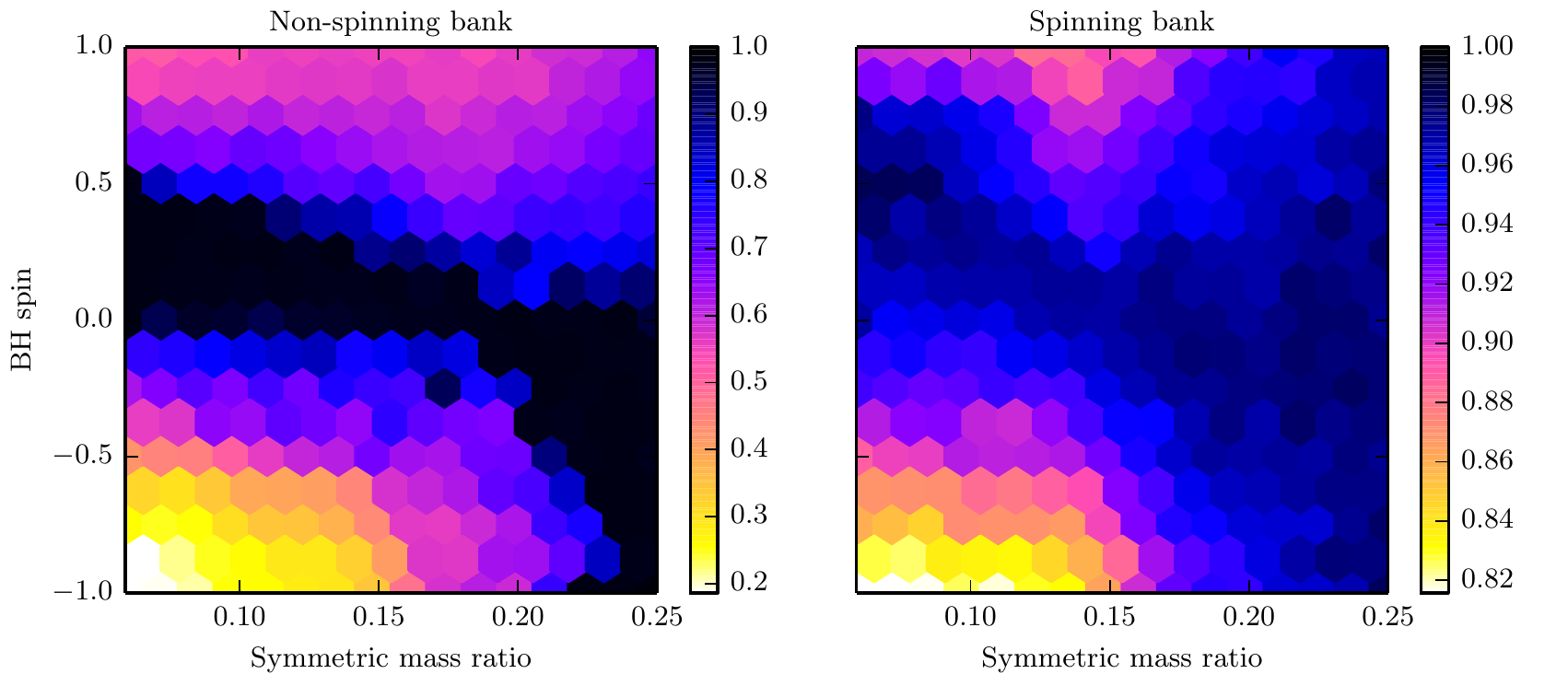}
    \caption{Performance of the two template banks across the whole
             non-precessing NS-BH parameter space. The color scale shows
             the lowest fitting factor found in each hexagonal bin (note the
             different color scales). The region of poor match in the
             spinning bank is likely due to the different
             termination conditions of templates and test waveforms.}
    \label{fig:overallbanksim}
\end{figure*}

As an overall test of the performance of the two banks over the NS-BH
parameter space, we calculate fitting factors with SpinTaylorT2 
signals uniformly distributed across the parameter space. The result 
is shown in Fig.~\ref{fig:overallbanksim},
illustrating the deficiency of the non-spinning bank over the 
parameter space.  It can be clearly seen that the values of $\chibh$ at 
which the match suddenly drops are a function of $\eta$; for equal-mass 
systems the match starts to drop for $\chibh \gtrsim 0$. Interestingly, 
the spinning bank can have a mismatch as large as 15\% in some parts of the 
parameter space. As noted already, this is likely an effect of the different
(mass- and spin-dependent) termination conditions of the template and test
waveforms. In fact, a template terminating before the signal loses the signal
power contained between \ac{ISCO} and \ac{MECO}, while in the case of a
template terminating after \ac{MECO} the \ac{SNR} normalization defined in
Eq.~\ref{eq:sigmasq} is too large; both cases result in an effective \ac{SNR}
loss. This hypothesis is supported by the fact that most of the residual
mismatch on the right plot of Fig.~\ref{fig:overallbanksim} covers the region
of $(\eta, \chibh)$ plane where a large difference exists between the \ac{MECO}
frequency and either the fixed termination at 1000\,Hz assumed in constructing
the bank or the \ac{ISCO} termination used for the template matched filtering.

Although our test signals and templates both use spinning phase corrections up
to 2.5 pN order, 3.5 pN corrections have been implemented during the
development of this paper and are now ready to be used in searches
\cite{Marsat:2012fn, PhysRevD.88.083002}. Unfortunately, including 3.5 pN
spinning terms in the \ac{MECO} definition can lead to very different
termination frequencies for our SpinTaylorT2 signals, introducing technical
difficulties which complicate our fitting factor calculation. Nevertheless, as
a rough characterization of the effect of 3.5 pN terms, we test our
non-spinning, 2.5 pN TaylorF2 bank against \ac{ISCO}-terminated TaylorF2
signals with 2.5 pN and 3.5 pN spinning terms. We find that the largest
variation in fitting factor when going from 2.5 to 3.5 pN signals is $\sim
0.05$, which is comparable to the maximum mismatch used for constructing the
bank and well below the loss due to neglecting spinning terms altogether. A
more detailed characterization of the inclusion of 3.5 pN spin terms represents
a separate study, but we see no reason for not using the best available phasing
in future searches.

\section{Details of the pipeline}
\label{sec:pipelinedetails}

After having described the template banks, the next step is evaluating the
performance of a full search pipeline running on realistic data.  This section
describes in detail the various components of the pipeline that we implement
(see Fig.~\ref{fig:pipeline}) and the corresponding parameter choices.

\subsection{Inspiral trigger generation}
\label{subsec:trig-gen}

Strain data are first processed by a PyCBC implementation of the standard
\texttt{FindChirp} algorithm \cite{Allen:2005fk} used in previous \ac{CBC} searches.
After conditioning the data \cite{Babak:2012zx} and estimating the noise
\ac{PSD}, the \ac{SNR} time series is computed for each template with a 
low-frequency cutoff of 30\,Hz. This is lower than past searches because
we are targeting the early advanced-LIGO sensitivity.  Local maxima of the 
\ac{SNR} time series that satisfy the condition $\rho > 5.9$ are identified. 
The $\chi_r^2$-statistic is then
computed for each surviving maximum via (\ref{eq:chisquare}) using 16
frequency bands, as is typical in \ac{CBC} searches \cite{Abbott:2009tt,
  oai:arXiv.org:1209.6533, Colaboration:2011np, Abadie:2011kd, Abadie:2010yb,
  Abbott:2009qj, PRD.73.102002, PRD.69.122001, PRD.72.082001, PRD.72.082002,
  0004-637X-715-2-1453}, and combined with the \ac{SNR} via (\ref{eq:newsnr}) to
obtain $\newsnr$. In order to reduce the very large number of maxima produced
by glitches, only those with $\newsnr > 5.9$ are kept as candidate triggers.
Such triggers are then stored in a MongoDB database \cite{mongodb}, where they
can be conveniently accessed by the next processing stages and also queried to
investigate the features of the data and the search.

The thresholds on \ac{SNR} and re-weighted \ac{SNR} used here are higher than
past \ac{CBC} searches (e.g.~\cite{Abbott:2009qj,Abadie:2010yb,Abadie:2011kd,
Colaboration:2011np,Babak:2012zx}) as they are chosen to fit the triggers
into the available database storage space, which is limited in our
prototype setup.  There is, in principle, no technical barrier to
extending the storage space of the database to handle a larger number
of events, which would allow one to lower the threshold back to the
usual value.

Even if the noise \ac{PSD} is continuously estimated from the data in order to
evaluate the \ac{SNR}, our template banks are constructed using the early
advanced-LIGO model \ac{PSD} and thus are constant for the whole data set. This
is also a notable difference with past searches, where template banks were
regenerated on a time scale of $\sim$ 30 minutes to account for the
variability of the noise \ac{PSD}.  Our choice is based on simplicity and the
relatively high computational cost of template bank generation. It is also
partly justified by the fact that the synthetic data we analyze is recolored to
the same noise curve used for constructing the banks.  Although the impact of a
fixed or varying bank on the sensitivity of a search is not yet fully
understood, we expect our choice to have a small effect on the result of our
comparison.

\subsection{Coincidence and clustering}
\label{subsec:coinc}

The next stage of the pipeline is the identification of triggers in
coincidence between the Hanford and Livingston detectors. Although the
recolored strain data we analyze covers a two-month period, the different
duty cycles of the detectors reduce the amount of data analyzed
in coincidence to about 25 days.

Because we store the triggers in a centralized database, different
coincidence methods can be applied to them. We choose an exact-match
method, where a trigger in detector A can only form a coincidence with
a trigger in detector B if the two triggers share the same template,
similar to \cite{Privitera:2013xza}.
This method has the advantage of simplicity and is straightforwardly 
applicable to parameter spaces of any dimensionality.
It requires however a common template bank for all detectors, another
difference with respect to past \ac{CBC} searches, as discussed above. 
A systematic comparison of different coincidence methods is outside
the scope of this paper.
Considering that the maximum arrival time delay between Hanford and
Livingston is $\sim 10$\,ms, and that the uncertainty in coalescence time
is of the order of a few milliseconds \cite{Nielsen:2012sb}, we choose a
conservative coincidence window of $\pm15$\,ms. Each pair of coincident
triggers is stored in the database and is tagged with the \emph{combined}
\ac{SNR} and re-weighted \ac{SNR}, defined respectively by summing in
quadrature the single-detector \acp{SNR} and re-weighted \acp{SNR}.

In order to keep only the most representative trigger among all the
triggers produced by a single inspiral signal or glitch, a final
clustering step is performed on coincident triggers. A trigger is
defined as representative if no other triggers with higher combined
re-weighted \ac{SNR} exist within $\pm 0.5$\,s.

Finally, in order to study the distribution of single-detector false
alarms, we also perform clustering of single-detector triggers.
This works in the same way as coincidence clustering, but it
uses a window of $\pm 15$\,ms only.

\subsection{Background and sensitivity estimation}
\label{subsec:bg}

In order to estimate the sensitivity of the search, we need to
determine i) the background rate of candidates in the absence of
astrophysical signals, and ii) how well the pipeline is able to
detect simulated NS-BH signals.

As usual in \ac{CBC} searches \cite{Colaboration:2011np}, we sample the
background distribution via time slides, i.e.\ by repeating the coincidence
step many times, each with a different time delay applied to triggers
from one of the detectors. We use 800 time delays, all multiples of 5
seconds. To avoid the possibility of true signals contaminating the
background, coincident triggers with zero time delay are excluded
from the sample.

We estimate the sensitivity to a population of NS-BH binaries by simulating
each binary's gravitational waveform, adding it to the strain data, analyzing
the data and recovering coincident triggers (if any) corresponding to each
coalescence. We perform three separate analysis runs with simulated signals
spaced over the full duration of recolored data at intervals of $\sim$ 10
minutes, resulting in $\sim 3\times10^4$ signals in total. The source
population is chosen to cover the parameter space reasonably broadly, while
being astrophysically plausible. The \ac{BH} mass is assigned a Gaussian
distribution centered on 7.8\,$\msun$ with a standard deviation of 3\,$\msun$,
truncated to the $[3,12]\,\msun$ range. The mean value is motivated by
\cite{Ozel:2010su} which suggests a mass distribution $(7.8 \pm 1.2)\,\msun$
for low-mass X-ray binaries; we choose a broader distribution with the same
mean.  The \ac{NS} mass is also Gaussian distributed with mean 1.35\,$\msun$
and standard deviation 0.13\,$\msun$ (following \cite{Kiziltan:2013oja}),
truncated to $[1,2]\,\msun$. Since this study ignores precession, both spins
are aligned with the orbital angular momentum. Both $\chibh$ and $\chins$ are
distributed uniformly, over ranges $[-0.99,0.99]$ and $[-0.05,0.05]$
respectively; as described later, however, we also consider three subsets of
the $\chibh$ range. The orbital angular momentum is distributed isotropically.

For the distance distribution, it is useful to introduce the notion of a
\emph{chirp distance}\footnote{See for instance \cite{Abbott:2009tt}, but note
that their definition uses the \emph{effective} rather than physical distance.}.
In the frequency domain, the amplitude of the signal in the restricted
post-Newtonian approximation is proportional to $\mathcal{M}^{5/6}/D$ with
$\mathcal{M}$ and $D$ being respectively the chirp mass and distance defined
in Sec.~\ref{sec:pipeline}.  The chirp distance is then defined as
\begin{equation}
    \mathcal{D} = D \left( \frac{\mathcal{M}_{\rm BNS}}{\mathcal{M}}\right)^{5/6}
\end{equation}
with $\mathcal{M}_{\rm BNS} \simeq 1.22$ $\msun$ being the chirp mass of a
canonical binary \ac{NS} system.  This quantity conveniently absorbs all the
mass dependent terms in the amplitude: to a first approximation, the detection
efficiency should have no additional mass dependence. We then simulate a
uniform distribution of sources over chirp distance, in the interval
$[1,160]$\,Mpc. Though unphysical, the choice of uniform chirp distance ensures
that i) the efficiency-vs-distance curve is sampled accurately across its
variation from 1 to 0 and ii) the most massive sources do not dominate the
recovered sample simply because of their high mass. As for our fitting-factor
calculations in Sec.~\ref{sec:templatebank}, the signal waveforms are simulated
via the standard SpinTaylorT2 approximant from LALSimulation \cite{LAL},
starting at 20 Hz and terminating at the \ac{MECO}.

The sensitivity of the searches is estimated by applying a window around 
the parameters of each source and recovering the most significant coincident
trigger within that window. Based on the results of the fitting-factor
simulations in Sec.~\ref{sec:templatebank}, we choose a coalescence-time
window of $\pm 0.5$\,s and a chirp-mass window of $\pm0.6$\,$\msun$. The
figure of merit we compute to compare the sensitivity of the two searches is
\begin{equation}
    V(\rho^*) = \frac{\sum_i \mathcal{D}^2_i P_i(\rho^*)}{\sum_i \mathcal{D}^2_i}
\end{equation}
where $P_i(\rho^*) = 1$ if source $i$ is recovered with a ranking statistic
larger than $\rho^*$ and equals 0 otherwise, and $\mathcal{D}_i$ is the chirp 
distance of source $i$. Here we use the quadrature sum of re-weighted SNRs 
$\newsnr$ over coincident triggers as ranking statistic. 
The $\mathcal{D}^2$ weighting corrects the figure of merit for
the unphysical distance distribution of the simulated binaries, such that
$V(\rho^*)$ is proportional to the sensitive volume of the search, which in 
turn is proportional to the expected rate of detections~\cite{Finn:1992xs}.

\section{Results}
\label{sec:results}

\subsection{Background}

Due to the increased dimensionality of the parameter space when going
from non-spinning to spinning templates, we expect a higher
false-alarm rate for the spinning search both in single-detector
triggers as well as in triggers coincident between the two detectors.

\begin{figure*}
    \centering
    \includegraphics[width=2\columnwidth]{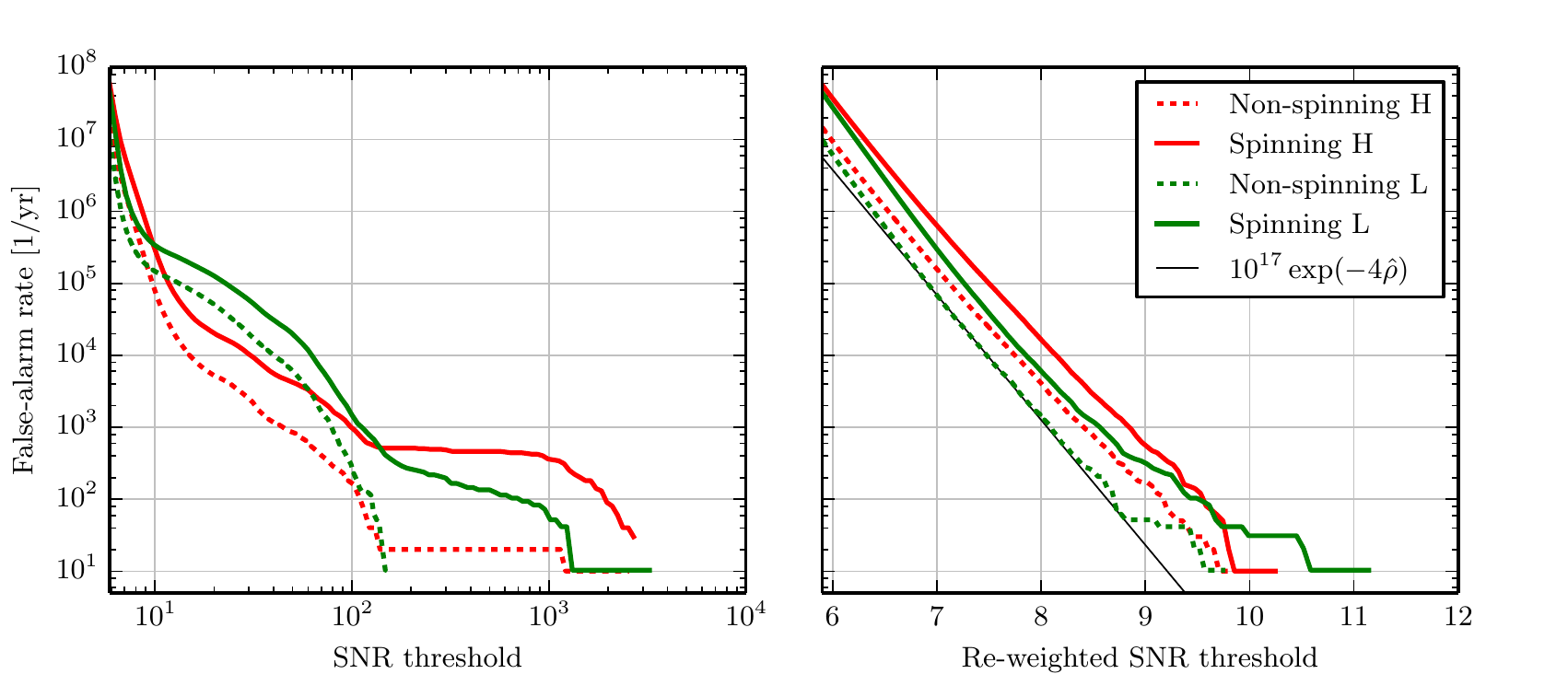}
    \caption{Rate of single-detector false alarms for the spinning and
             non-spinning searches as a function of the threshold on
             \ac{SNR} and re-weighted \ac{SNR}.}
    \label{fig:background-single}
\end{figure*}

Single-detector background triggers associated with \ac{SNR} and
re-weighted \ac{SNR} are shown in Fig.~\ref{fig:background-single}.
The spinning search clearly has a higher false-alarm rate for both
detection statistics. As is well known from past \ac{CBC} searches
\cite{Blackburn:2008ah}, the \ac{SNR} background exhibits a large
tail associated with non-Gaussian transient glitches\footnote{Note that
in our test we analyze all available science-mode data, including a few
poor-quality data segments which a real search would exclude via 
data-quality flags \cite{Slutsky:2010ff,Babak:2012zx}. Thus, the tail
in our \ac{SNR} background is likely exaggerated.}. The spinning
search seems to be affected more by glitches, as can be seen from the
much larger tail at high \ac{SNR}. Thanks to the effectiveness of the
$\chi^2$ test, however, the re-weighted \ac{SNR} is almost tail-free,
although we find that strong glitches can still lead to false
alarms noticeably stronger than what is typical in stationary Gaussian
noise. The increase of false-alarm rate associated with the re-weighted
\ac{SNR} is proportional to the increase in number of templates ($\sim 5
\times$) for almost all values of the threshold. Applying a fixed
threshold in false-alarm rate implies an increase in single-detector
re-weighted \ac{SNR} of 0.5 or less when going from the non-spinning to
the spinning search. The background distribution of
re-weighted \ac{SNR} falls approximately like $\exp(-k\hat{\rho})$ with
$k\sim 4$, such that if the total rate of triggers increases by a
factor $\alpha$, the increase in statistic threshold required to
compensate this increase (and thus preserve the same false-alarm rate)
is only $\Delta \hat{\rho} \sim \log(\alpha)/k$.

\begin{figure}
    \centering
    \includegraphics[width=\columnwidth]{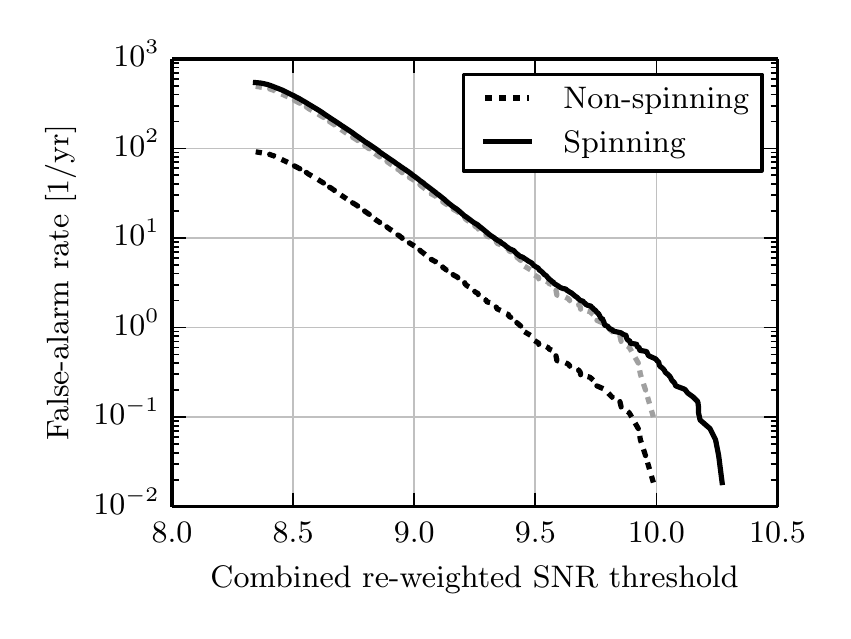}
    \caption{Rate of coincident false alarms for the spinning and non-spinning
             searches as a function of the threshold on combined re-weighted
             \ac{SNR}. The shaded dotted curve shows the non-spinning curve
             multiplied by the relative number of templates of the two
             searches ($\sim 5 \times$).}
    \label{fig:background-coinc}
\end{figure}

The coincident background distribution over the combined (quadrature sum) 
$\hat{\rho}$ statistic is shown in Fig.~\ref{fig:background-coinc}.
As for single-detector backgrounds, the larger false-alarm rate of the
spinning search is consistent with the increase in template bank size
except at very low rate, where our background sample is likely affected
by a small number of loud glitches. Nevertheless, the increase in
ranking statistic required to maintain a fixed false-alarm rate from
non-spinning to spinning search is only about 0.3.

\subsection{Signal recovery and sensitivity}

\begin{figure*}
 \centering
 \includegraphics[width=2\columnwidth]{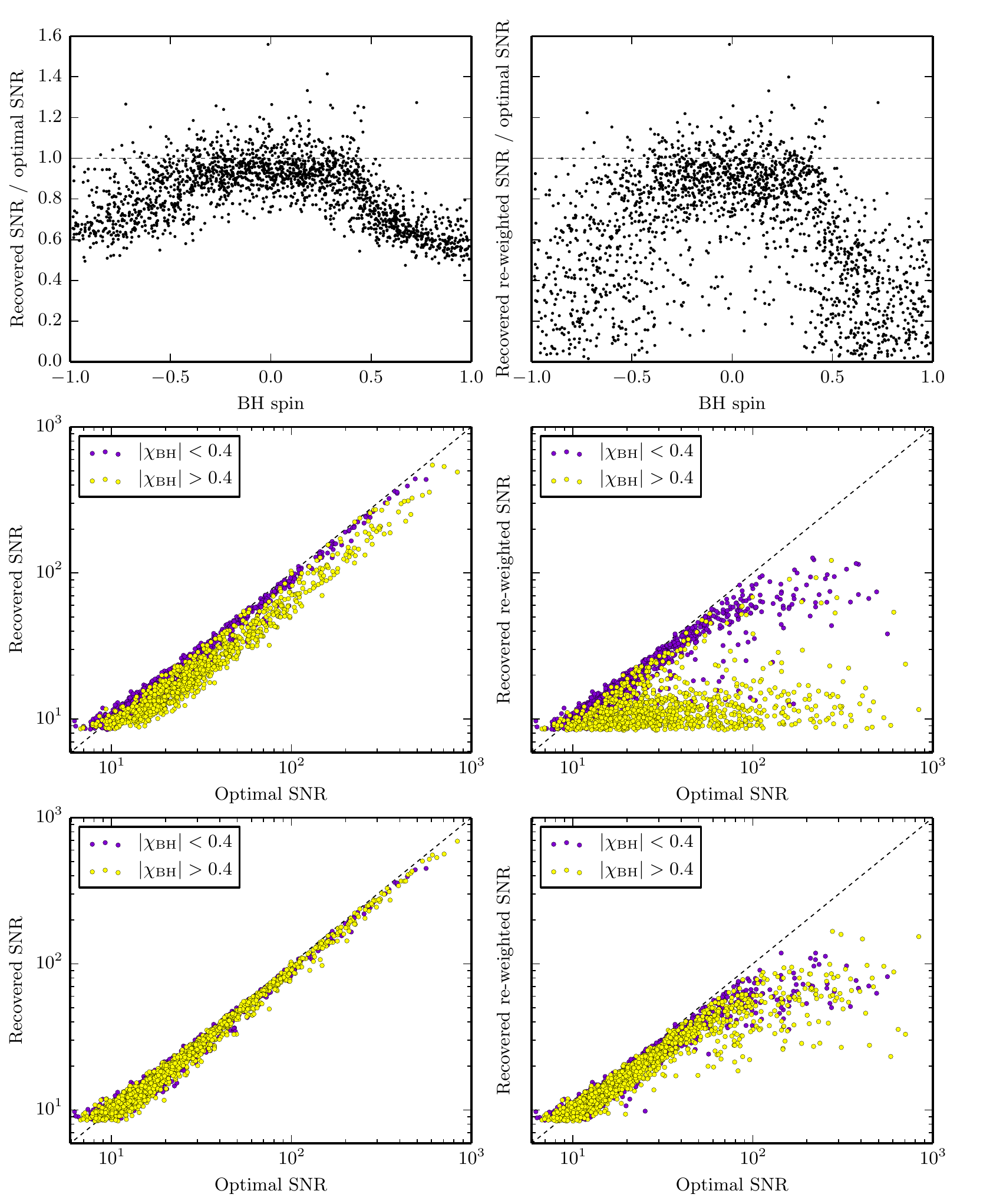}
 \caption{Top row: fraction of optimal combined \ac{SNR} and re-weighted
 \ac{SNR} recovered by the non-spinning search for each simulated (and found)
 source, as a function of the \ac{BH} spin (compare with
 Fig.~\ref{fig:nospinmatches}). Middle row: combined \ac{SNR} and re-weighted
 \ac{SNR} recovered by the non-spinning search for each found source versus the
 optimal combined \ac{SNR}. The color distinguishes between high ($>0.4$) and
 low ($<0.4$) \ac{BH} spin magnitude.  Bottom row: same as middle row, for the
 spinning search.}
 \label{fig:recovery}
\end{figure*}

As a check of the correct behavior of the search, we first calculate the
\emph{optimal} \ac{SNR} of each simulated source, i.e.\ the \ac{SNR} obtained
for a vanishing noise realization and a perfectly matched template. For each
source producing a coincidence in both searches, we compare its
combined optimal \ac{SNR} with the combined \ac{SNR} and re-weighted \ac{SNR}
actually recovered by the searches (the $\chi^2$ value for a zero noise
realization and ideal template is also zero, thus the optimal re-weighted
\ac{SNR} is equal to the optimal \ac{SNR}). We find that the non-spinning
search fails to recover a noticeable fraction of both \ac{SNR} and re-weighted
\ac{SNR} for $|\chibh| \gtrsim 0.5$, which is roughly consistent with the
fitting factor calculations (Fig.~\ref{fig:recovery}, top and middle rows).
The impact of the $\chi^2$ test on spinning signals is particularly dramatic,
as the loss in re-weighted \ac{SNR} is much larger than the loss in \ac{SNR}.
The spinning search, instead, recovers the expected \ac{SNR} almost completely,
for all values of the \ac{BH} spin (Fig.~\ref{fig:recovery}, bottom-left plot).
Note however that sources with optimal \ac{SNR} larger than $\sim 100$ have a
significant loss in re-weighted \ac{SNR} even in the spinning bank; in fact,
the re-weighted \ac{SNR} appears to asymptote to a finite value when the
optimal \ac{SNR} becomes very large (Fig.~\ref{fig:recovery}, bottom-right
plot). This can be explained by the small but non-zero residual mismatch which
is also present in the spinning bank. In fact, with any non-zero mismatch, at
some (large) value of $\rho$ the $\chi^2$ statistic eventually starts growing
like $\rho^2$ \cite{Allen:2005fk}. Combining this fact with the definition of
re-weighted \ac{SNR} (Eq.~\ref{eq:newsnr}) results in a finite re-weighted
\ac{SNR} for arbitrarily large \ac{SNR}.

\begin{figure*}
    \centering
    \includegraphics[width=2\columnwidth]{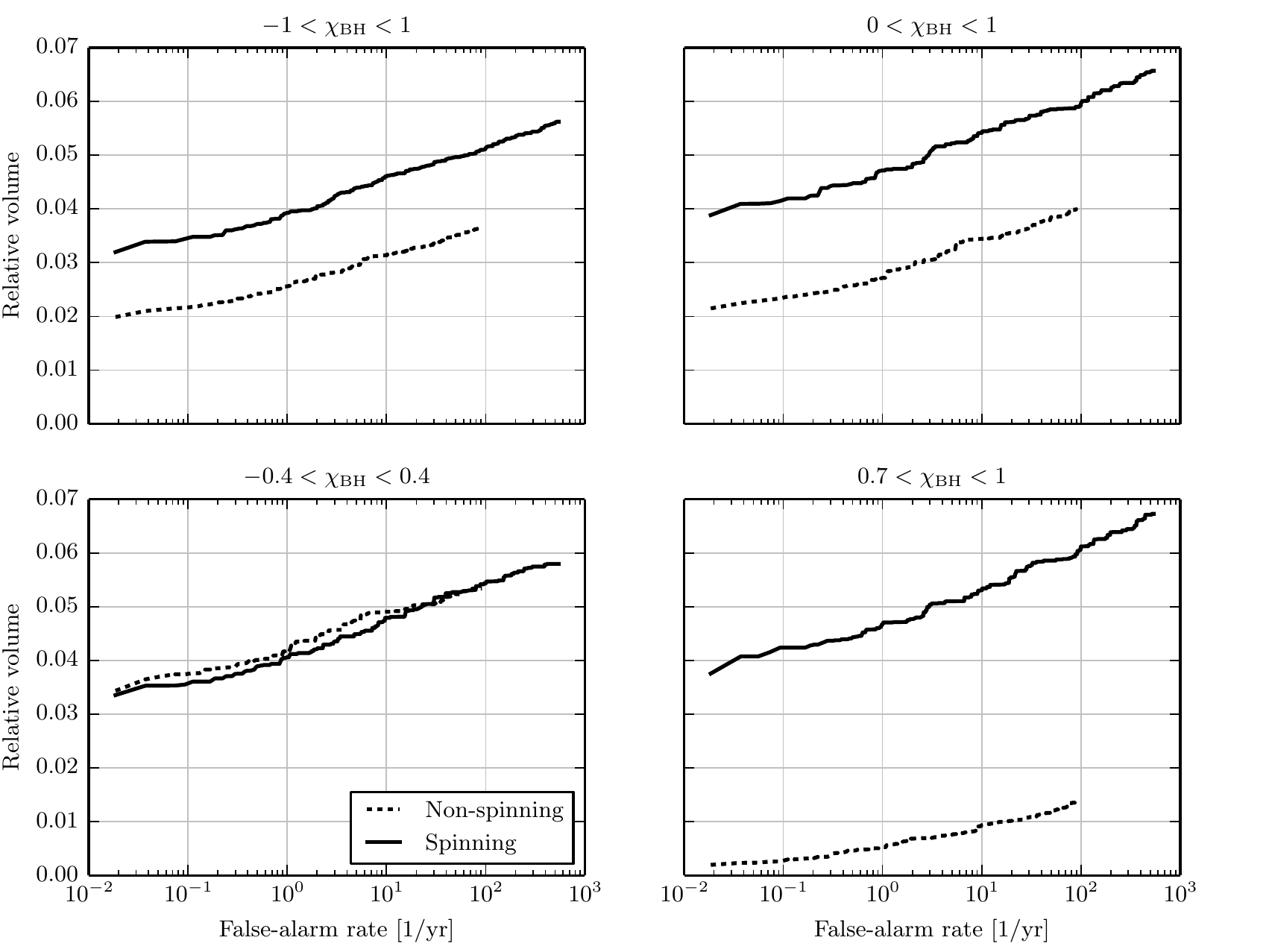}
    \caption{\ac{ROC} curves for the spinning and non-spinning searches, comparing
    the relative sensitive volume (or number of detections) at fixed
    false-alarm rate. The four panels assume NS-BH systems with different
    limits on a uniform distribution of \ac{BH} spins.}
    \label{fig:rocs}
\end{figure*}

Considering the relative sensitivity of the two searches at fixed false-alarm
rate, we find that it depends strongly on the distribution of \ac{BH} spins.
Fig.~\ref{fig:rocs} shows the \ac{ROC} curves for four populations of NS-BH binaries
associated with different \ac{BH} spin distributions. As done throughout this paper,
all cases assume alignment between the \ac{BH} spin and the orbital angular momentum.

Assuming \acp{BH} can have any spin magnitude within the limits of the Kerr
bound, we obtain an increase in sensitivity of the spinning search between 40\%
and 60\% depending on the false-alarm rate. A slightly larger improvement is
obtained if $\chibh$ is restricted to be positive. If all \acp{BH} are highly
spinning and positively aligned with the orbital angular momentum, however, the
spinning search can be $\mathcal{O}(10)$ times more sensitive than the
non-spinning one at interesting false-alarm rates. This large difference can be
understood by considering the dramatic loss in re-weighted \ac{SNR} of the
non-spinning search, which is due in part to the large \ac{SNR} loss and in
part to the poor $\chi^2$ value of highly-spinning signals. If the search could
be carried out using the standard \ac{SNR} as the ranking statistic, or if we
tuned the $\chi^2$ veto differently, the improvement could be significantly
less dramatic, but likely still interesting; Fig.~\ref{fig:overallbanksim}
(left plot) and Fig.~\ref{fig:recovery} (top-left plot) both suggest a factor
of $0.6^{-3}\simeq 4.6$ when using the \ac{SNR} as the ranking statistic. We
also note that the improvement could be less dramatic if precession is included
in the simulated binaries, but this will be studied in a forthcoming paper. For
weakly spinning \acp{BH} ($|\chibh|<0.4$), the spinning search is a few percent
less sensitive, as can be expected from the larger background, although the
difference is comparable with the statistical fluctuations of our \ac{ROC}
curves.

Our signals provide insufficient statistics for studying the case of very
small \ac{BH} spins. Nevertheless, we can conclude with a back-of-the-envelope
comparison of the searches assuming a worst-case population of exactly
non-spinning \acp{BH}. Using the background curves from
Fig.~\ref{fig:background-coinc} and assuming a non-spinning search with
detection threshold $\rho^{*} \gtrsim 9.5$, the relative sensitive volume of
the spinning search would be
\begin{equation}
    V(\rho^{*}) \simeq \left( \frac{\rho^{*}}{\rho^{*} + 0.3} \right)^3 > 90\%.
\end{equation}
In the worst case, therefore, the spinning search would lose 10\% or less of
the signals; the major burden would be the larger computational cost.

\section{Conclusions}
\label{sec:conclusions}

We show for the first time how an aligned-spin search for NS-BH binaries can
be successfully implemented in the advanced detector era. We demonstrate a
prototype search in the PyCBC framework which contains
all the essential elements of a realistic \ac{CBC} search: matched
filtering with a template bank, signal-based vetoes, a suitable
ranking statistic, coincidence, clustering, background estimates using
time slides and sensitivity estimates via a simulated population of signals.
The use of re-weighted \ac{SNR} as event ranking statistic
\cite{Colaboration:2011np} is sufficient to reduce the background to a level
that makes the search more sensitive than traditional non-spinning
searches over the full range of \ac{BH} spins.

An important element in making this work is running the analysis on \acp{GPU} using
PyCBC. The design of PyCBC and the available tools make it easy to put
together such a pipeline, and the use of \acp{GPU} speeds up the analysis
relative to using \acp{CPU}. Such an analysis would have been much harder
with earlier technology.

Our spinning search has improved sensitivity relative to a traditional
non-spinning search. The improvement in sensitivity depends strongly on the
exact distribution of \ac{BH} spins.  If the magnitude of the dimensionless
\ac{BH} spins (taken to be parallel to the orbital angular momentum) is mostly
below $\sim 0.4$, the spinning and non-spinning searches have approximately the
same sensitivity, despite the fact that the spinning template bank contains
many more templates.  If the \ac{BH} spin is distributed uniformly in the range
$(-1,1)$, then the spinning search has approximately $50\%$ greater
astrophysical reach as measured by sensitive volume. The increase is one order
of magnitude if the spin is uniformly distributed in the interval $(0.7,1)$.
We stress that these results assume that i) all systems are
non-precessing, and ii) the \ac{BH} and \ac{NS} mass distributions are
Gaussians with means $7.8\,\msun$ and $1.35\,\msun$ and standard deviations
$3\,\msun$ and $0.13\,\msun$ respectively, with the additional constraints
$\mbh \in [3,12]\,\msun$ and $\mns \in [1,2]\,\msun$.
Although a careful study of the effect of different mass distributions is
outside the scope of this paper, different distributions are unlikely to change
the fact that a spinning search is more sensitive than a non-spinning search;
this would require an unrealistic distribution restricted to the
parameter-space region where the non-spinning bank performs well (see
Fig.~\ref{fig:overallbanksim}).

The available X-ray data and population synthesis studies suggest that
the spin parameters of \acp{BH} may be reasonably large, greater
than $\sim 0.7$ in the mass range we used. If this is the case, then
for aligned systems an aligned spin search offers a significant
improvement in sensitive volume and hence event rate, relative to traditional
non-spinning searches.  The improvement in search sensitivity could then
mean the difference between detection and non-detection, depending on
the astrophysical rate of NS-BH coalescence events.

We base our conclusions on an idealized noise \ac{PSD} which could represent
the early runs of advanced LIGO and we employ template banks fixed in time
and identical between different detectors. We also show that the \ac{SNR} loss
of a non-spinning bank relative to a spinning one depends on the noise
\ac{PSD}.  If advanced LIGO's sensitivity has a significantly smaller bandwidth
than our model, or if its noise \ac{PSD} turns out to have a large variability
over a time scale of a few months, the sensitivity of a search to spin effects
could be smaller and thus our spinning search could be less beneficial. We
argue however that these are unlikely scenarios, as the evolution of advanced
LIGO will drive towards the large bandwidth of the final design sensitivity.
In addition, if the bandwidth is so narrow that spin effects are less
important than we find here, the spinning search would be at least as sensitive
as the non-spinning one for a uniform distribution of \ac{BH} spins and thus
would only produce a larger computational cost.

The search methodology we present is straightforward and based on previous
\ac{CBC} analyses, but it has not been fully optimized. We expect that
further improvements to the analysis will be possible. These include: i)
constructing a better template bank using the full 3.5 pN phasing, a geometric
placement algorithm and a mass- and spin-dependent upper frequency cutoff;
ii) correcting the event ranking statistic to reflect the non-uniform
distribution of templates over component masses and spins, as described in
\cite{PhysRevD.89.062002}; iii) improving the ranking statistic by
accounting for event distributions over extrinsic parameters such as
coalescence time and amplitude; and iv) tuning the coincidence and
clustering steps.
Including merger and ringdown effects should further improve the search
sensitivity at the higher mass end
of the parameter space. The impact of poor data quality on the
computing time of inspiral jobs deserves further research. Improved data
conditioning techniques such as gating, i.e.\ appropriately windowing
out the data segment in the vicinity of high-amplitude glitches, are
under investigation. In developing and testing such improvements to the
search, PyCBC will be an essential tool.  Finally, investigating the
effects of precession will be important as well.  This search should
be seen as an intermediate step towards a full precessing search; a
study is underway to quantify how well the current pipeline performs in
detecting precessing signals.

Our conclusions remain robust towards further tuning: when using a template
bank that includes the effect of spin, with a signal-based veto such as
$\chi^2$ and the infrastructure required to run a search to completion, the
gain in signal sensitivity easily outweighs the increase in background.
Thus we advise an aligned-spin search rather than a non-spinning search for
NS-BH binaries, even for the early advanced detectors.

\begin{acknowledgments}
  We thank Gergely Debreczeni, Mate Ferenc Nagy, Frank Ohme, Gianluca Guidi
  and Richard O'Shaughnessy for useful discussions and comments, and
  Carsten Aulbert and Oliver Bock for valuable support in using the Atlas
  cluster and MongoDB at the Albert Einstein Institute.  We also thank the
  LIGO collaboration for providing the recolored synthetic strain data we
  analyzed.  TDC is supported by the International Max-Planck Research
  School on Gravitational-Wave Astronomy.  AJM and JLW are supported
  in part by the Pursuit program and Office of Undergraduate Research
  of ACU.  This work is supported by National Science Foundation
  awards PHY-0847611, PHY-0854812, and a Cottrell Scholar award from
  the Research Corporation for Science Advancement. Part of the
  computations used in this work were performed on the Syracuse
  University Gravitation and Relativity cluster, which is supported by
  NSF awards PHY-1040231 and PHY-1104371.

  This paper has LIGO document number LIGO-P1400053.
\end{acknowledgments}

\bibliography{pycbcpaper}{}

\begin{thebibliography}{93}%
\makeatletter
\providecommand \@ifxundefined [1]{%
 \@ifx{#1\undefined}
}%
\providecommand \@ifnum [1]{%
 \ifnum #1\expandafter \@firstoftwo
 \else \expandafter \@secondoftwo
 \fi
}%
\providecommand \@ifx [1]{%
 \ifx #1\expandafter \@firstoftwo
 \else \expandafter \@secondoftwo
 \fi
}%
\providecommand \natexlab [1]{#1}%
\providecommand \enquote  [1]{``#1''}%
\providecommand \bibnamefont  [1]{#1}%
\providecommand \bibfnamefont [1]{#1}%
\providecommand \citenamefont [1]{#1}%
\providecommand \href@noop [0]{\@secondoftwo}%
\providecommand \href [0]{\begingroup \@sanitize@url \@href}%
\providecommand \@href[1]{\@@startlink{#1}\@@href}%
\providecommand \@@href[1]{\endgroup#1\@@endlink}%
\providecommand \@sanitize@url [0]{\catcode `\\12\catcode `\$12\catcode
  `\&12\catcode `\#12\catcode `\^12\catcode `\_12\catcode `\%12\relax}%
\providecommand \@@startlink[1]{}%
\providecommand \@@endlink[0]{}%
\providecommand \url  [0]{\begingroup\@sanitize@url \@url }%
\providecommand \@url [1]{\endgroup\@href {#1}{\urlprefix }}%
\providecommand \urlprefix  [0]{URL }%
\providecommand \Eprint [0]{\href }%
\providecommand \doibase [0]{http://dx.doi.org/}%
\providecommand \selectlanguage [0]{\@gobble}%
\providecommand \bibinfo  [0]{\@secondoftwo}%
\providecommand \bibfield  [0]{\@secondoftwo}%
\providecommand \translation [1]{[#1]}%
\providecommand \BibitemOpen [0]{}%
\providecommand \bibitemStop [0]{}%
\providecommand \bibitemNoStop [0]{.\EOS\space}%
\providecommand \EOS [0]{\spacefactor3000\relax}%
\providecommand \BibitemShut  [1]{\csname bibitem#1\endcsname}%
\let\auto@bib@innerbib\@empty
\bibitem [{\citenamefont {Harry}(2010)}]{Harry:2010zz}%
  \BibitemOpen
  \bibfield  {author} {\bibinfo {author} {\bibfnamefont {G.~M.}\ \bibnamefont
  {Harry}} (\bibinfo {collaboration} {LIGO Scientific Collaboration}),\ }\href
  {\doibase 10.1088/0264-9381/27/8/084006} {\bibfield  {journal} {\bibinfo
  {journal} {Class.Quant.Grav.}\ }\textbf {\bibinfo {volume} {27}},\ \bibinfo
  {pages} {084006} (\bibinfo {year} {2010})}\BibitemShut {NoStop}%
\bibitem [{\citenamefont {Accadia}\ \emph {et~al.}(2012)\citenamefont {Accadia}
  \emph {et~al.}}]{Accadia:2012zzb}%
  \BibitemOpen
  \bibfield  {author} {\bibinfo {author} {\bibfnamefont {T.}~\bibnamefont
  {Accadia}} \emph {et~al.} (\bibinfo {collaboration} {VIRGO Collaboration}),\
  }\href {\doibase 10.1088/1748-0221/7/03/P03012} {\bibfield  {journal}
  {\bibinfo  {journal} {JINST}\ }\textbf {\bibinfo {volume} {7}},\ \bibinfo
  {pages} {P03012} (\bibinfo {year} {2012})}\BibitemShut {NoStop}%
\bibitem [{\citenamefont {Somiya}(2012)}]{Somiya:2011np}%
  \BibitemOpen
  \bibfield  {author} {\bibinfo {author} {\bibfnamefont {K.}~\bibnamefont
  {Somiya}} (\bibinfo {collaboration} {KAGRA Collaboration}),\ }\href {\doibase
  10.1088/0264-9381/29/12/124007} {\bibfield  {journal} {\bibinfo  {journal}
  {Class.Quant.Grav.}\ }\textbf {\bibinfo {volume} {29}},\ \bibinfo {pages}
  {124007} (\bibinfo {year} {2012})},\ \Eprint {http://arxiv.org/abs/1111.7185}
  {arXiv:1111.7185 [gr-qc]} \BibitemShut {NoStop}%
\bibitem [{\citenamefont {Iyer}\ \emph {et~al.}(2011)\citenamefont {Iyer},
  \citenamefont {Souradeep}, \citenamefont {Unnikrishnan}, \citenamefont
  {Dhurandhar}, \citenamefont {Raja},\ and\ \citenamefont
  {Sengupta}}]{LIGOIndia}%
  \BibitemOpen
  \bibfield  {author} {\bibinfo {author} {\bibfnamefont {B.}~\bibnamefont
  {Iyer}}, \bibinfo {author} {\bibfnamefont {T.}~\bibnamefont {Souradeep}},
  \bibinfo {author} {\bibfnamefont {C.}~\bibnamefont {Unnikrishnan}}, \bibinfo
  {author} {\bibfnamefont {S.}~\bibnamefont {Dhurandhar}}, \bibinfo {author}
  {\bibfnamefont {S.}~\bibnamefont {Raja}}, \ and\ \bibinfo {author}
  {\bibfnamefont {A.}~\bibnamefont {Sengupta}},\ }\href@noop {} {\enquote
  {\bibinfo {title} {Ligo-india, proposal of the consortium for indian
  initiative in gravitational-wave observations (indigo)},}\ }\bibinfo
  {howpublished}
  {\url{https://dcc.ligo.org/cgi-bin/DocDB/ShowDocument?docid=75988}} (\bibinfo
  {year} {2011})\BibitemShut {NoStop}%
\bibitem [{\citenamefont {Abadie}\ \emph
  {et~al.}(2010{\natexlab{a}})\citenamefont {Abadie} \emph
  {et~al.}}]{Abadie:2010cf}%
  \BibitemOpen
  \bibfield  {author} {\bibinfo {author} {\bibfnamefont {J.}~\bibnamefont
  {Abadie}} \emph {et~al.} (\bibinfo {collaboration} {LIGO Scientific
  Collaboration, Virgo Collaboration}),\ }\href {\doibase
  10.1088/0264-9381/27/17/173001} {\bibfield  {journal} {\bibinfo  {journal}
  {Class.Quant.Grav.}\ }\textbf {\bibinfo {volume} {27}},\ \bibinfo {pages}
  {173001} (\bibinfo {year} {2010}{\natexlab{a}})},\ \Eprint
  {http://arxiv.org/abs/1003.2480} {arXiv:1003.2480 [astro-ph.HE]} \BibitemShut
  {NoStop}%
\bibitem [{\citenamefont {Abbott}\ \emph {et~al.}(2008)\citenamefont {Abbott}
  \emph {et~al.}}]{Abbott:2007ai}%
  \BibitemOpen
  \bibfield  {author} {\bibinfo {author} {\bibfnamefont {B.}~\bibnamefont
  {Abbott}} \emph {et~al.} (\bibinfo {collaboration} {LIGO Scientific
  Collaboration}),\ }\href {\doibase 10.1103/PhysRevD.78.042002} {\bibfield
  {journal} {\bibinfo  {journal} {Phys. Rev.}\ }\textbf {\bibinfo {volume}
  {D78}},\ \bibinfo {pages} {042002} (\bibinfo {year} {2008})},\ \Eprint
  {http://arxiv.org/abs/0712.2050} {arXiv:0712.2050 [gr-qc]} \BibitemShut
  {NoStop}%
\bibitem [{\citenamefont {Aasi}\ \emph
  {et~al.}(2013{\natexlab{a}})\citenamefont {Aasi} \emph
  {et~al.}}]{oai:arXiv.org:1209.6533}%
  \BibitemOpen
  \bibfield  {author} {\bibinfo {author} {\bibfnamefont {J.}~\bibnamefont
  {Aasi}} \emph {et~al.} (\bibinfo {collaboration} {LIGO Scientific
  Collaboration, Virgo Collaboration}),\ }\href {\doibase
  10.1103/PhysRevD.87.022002} {\bibfield  {journal} {\bibinfo  {journal} {Phys.
  Rev.}\ }\textbf {\bibinfo {volume} {D87}},\ \bibinfo {pages} {022002}
  (\bibinfo {year} {2013}{\natexlab{a}})},\ \Eprint
  {http://arxiv.org/abs/1209.6533} {arXiv:1209.6533 [gr-qc]} \BibitemShut
  {NoStop}%
\bibitem [{\citenamefont {Van Den~Broeck}\ \emph {et~al.}(2009)\citenamefont
  {Van Den~Broeck}, \citenamefont {Brown}, \citenamefont {Cokelaer},
  \citenamefont {Harry}, \citenamefont {Jones} \emph
  {et~al.}}]{VanDenBroeck:2009gd}%
  \BibitemOpen
  \bibfield  {author} {\bibinfo {author} {\bibfnamefont {C.}~\bibnamefont {Van
  Den~Broeck}}, \bibinfo {author} {\bibfnamefont {D.~A.}\ \bibnamefont
  {Brown}}, \bibinfo {author} {\bibfnamefont {T.}~\bibnamefont {Cokelaer}},
  \bibinfo {author} {\bibfnamefont {I.}~\bibnamefont {Harry}}, \bibinfo
  {author} {\bibfnamefont {G.}~\bibnamefont {Jones}},  \emph {et~al.},\ }\href
  {\doibase 10.1103/PhysRevD.80.024009} {\bibfield  {journal} {\bibinfo
  {journal} {Phys. Rev.}\ }\textbf {\bibinfo {volume} {D80}},\ \bibinfo {pages}
  {024009} (\bibinfo {year} {2009})},\ \Eprint {http://arxiv.org/abs/0904.1715}
  {arXiv:0904.1715 [gr-qc]} \BibitemShut {NoStop}%
\bibitem [{\citenamefont {Privitera}\ \emph {et~al.}(2014)\citenamefont
  {Privitera}, \citenamefont {Mohapatra}, \citenamefont {Ajith}, \citenamefont
  {Cannon}, \citenamefont {Fotopoulos} \emph {et~al.}}]{Privitera:2013xza}%
  \BibitemOpen
  \bibfield  {author} {\bibinfo {author} {\bibfnamefont {S.}~\bibnamefont
  {Privitera}}, \bibinfo {author} {\bibfnamefont {S.~R.~P.}\ \bibnamefont
  {Mohapatra}}, \bibinfo {author} {\bibfnamefont {P.}~\bibnamefont {Ajith}},
  \bibinfo {author} {\bibfnamefont {K.}~\bibnamefont {Cannon}}, \bibinfo
  {author} {\bibfnamefont {N.}~\bibnamefont {Fotopoulos}},  \emph {et~al.},\
  }\href {\doibase 10.1103/PhysRevD.89.024003} {\bibfield  {journal} {\bibinfo
  {journal} {Phys. Rev.}\ }\textbf {\bibinfo {volume} {D89}},\ \bibinfo {pages}
  {024003} (\bibinfo {year} {2014})},\ \Eprint {http://arxiv.org/abs/1310.5633}
  {arXiv:1310.5633 [gr-qc]} \BibitemShut {NoStop}%
\bibitem [{\citenamefont {Brown}\ \emph {et~al.}(2012)\citenamefont {Brown},
  \citenamefont {Harry}, \citenamefont {Lundgren},\ and\ \citenamefont
  {Nitz}}]{Brown:2012qf}%
  \BibitemOpen
  \bibfield  {author} {\bibinfo {author} {\bibfnamefont {D.~A.}\ \bibnamefont
  {Brown}}, \bibinfo {author} {\bibfnamefont {I.}~\bibnamefont {Harry}},
  \bibinfo {author} {\bibfnamefont {A.}~\bibnamefont {Lundgren}}, \ and\
  \bibinfo {author} {\bibfnamefont {A.~H.}\ \bibnamefont {Nitz}},\ }\href
  {\doibase 10.1103/PhysRevD.86.084017} {\bibfield  {journal} {\bibinfo
  {journal} {Phys. Rev.}\ }\textbf {\bibinfo {volume} {D86}},\ \bibinfo {pages}
  {084017} (\bibinfo {year} {2012})},\ \Eprint {http://arxiv.org/abs/1207.6406}
  {arXiv:1207.6406 [gr-qc]} \BibitemShut {NoStop}%
\bibitem [{\citenamefont {Harry}\ \emph {et~al.}(2013)\citenamefont {Harry},
  \citenamefont {Nitz}, \citenamefont {Brown}, \citenamefont {Lundgren},
  \citenamefont {Ochsner} \emph {et~al.}}]{Harry:2013tca}%
  \BibitemOpen
  \bibfield  {author} {\bibinfo {author} {\bibfnamefont {I.}~\bibnamefont
  {Harry}}, \bibinfo {author} {\bibfnamefont {A.}~\bibnamefont {Nitz}},
  \bibinfo {author} {\bibfnamefont {D.~A.}\ \bibnamefont {Brown}}, \bibinfo
  {author} {\bibfnamefont {A.}~\bibnamefont {Lundgren}}, \bibinfo {author}
  {\bibfnamefont {E.}~\bibnamefont {Ochsner}},  \emph {et~al.},\ }\href@noop {}
  {\  (\bibinfo {year} {2013})},\ \Eprint {http://arxiv.org/abs/1307.3562}
  {arXiv:1307.3562 [gr-qc]} \BibitemShut {NoStop}%
\bibitem [{\citenamefont {Mandel}\ and\ \citenamefont
  {O'Shaughnessy}(2010)}]{Mandel:2009nx}%
  \BibitemOpen
  \bibfield  {author} {\bibinfo {author} {\bibfnamefont {I.}~\bibnamefont
  {Mandel}}\ and\ \bibinfo {author} {\bibfnamefont {R.}~\bibnamefont
  {O'Shaughnessy}},\ }\href {\doibase 10.1088/0264-9381/27/11/114007}
  {\bibfield  {journal} {\bibinfo  {journal} {Class.Quant.Grav.}\ }\textbf
  {\bibinfo {volume} {27}},\ \bibinfo {pages} {114007} (\bibinfo {year}
  {2010})},\ \Eprint {http://arxiv.org/abs/0912.1074} {arXiv:0912.1074
  [astro-ph.HE]} \BibitemShut {NoStop}%
\bibitem [{\citenamefont {O'Shaughnessy}\ \emph {et~al.}(2005)\citenamefont
  {O'Shaughnessy}, \citenamefont {Kaplan}, \citenamefont {Kalogera},\ and\
  \citenamefont {Belczynski}}]{O'Shaughnessy:2005qc}%
  \BibitemOpen
  \bibfield  {author} {\bibinfo {author} {\bibfnamefont {R.~W.}\ \bibnamefont
  {O'Shaughnessy}}, \bibinfo {author} {\bibfnamefont {J.}~\bibnamefont
  {Kaplan}}, \bibinfo {author} {\bibfnamefont {V.}~\bibnamefont {Kalogera}}, \
  and\ \bibinfo {author} {\bibfnamefont {K.}~\bibnamefont {Belczynski}},\
  }\href {\doibase 10.1086/444346} {\bibfield  {journal} {\bibinfo  {journal}
  {Astrophys.J.}\ }\textbf {\bibinfo {volume} {632}},\ \bibinfo {pages} {1035}
  (\bibinfo {year} {2005})},\ \Eprint {http://arxiv.org/abs/astro-ph/0503219}
  {arXiv:astro-ph/0503219 [astro-ph]} \BibitemShut {NoStop}%
\bibitem [{\citenamefont {{Eichler}}\ \emph {et~al.}(1989)\citenamefont
  {{Eichler}}, \citenamefont {{Livio}}, \citenamefont {{Piran}},\ and\
  \citenamefont {{Schramm}}}]{1989Natur.340..126E}%
  \BibitemOpen
  \bibfield  {author} {\bibinfo {author} {\bibfnamefont {D.}~\bibnamefont
  {{Eichler}}}, \bibinfo {author} {\bibfnamefont {M.}~\bibnamefont {{Livio}}},
  \bibinfo {author} {\bibfnamefont {T.}~\bibnamefont {{Piran}}}, \ and\
  \bibinfo {author} {\bibfnamefont {D.~N.}\ \bibnamefont {{Schramm}}},\ }\href
  {\doibase 10.1038/340126a0} {\bibfield  {journal} {\bibinfo  {journal}
  {\nat}\ }\textbf {\bibinfo {volume} {340}},\ \bibinfo {pages} {126} (\bibinfo
  {year} {1989})}\BibitemShut {NoStop}%
\bibitem [{\citenamefont {{Narayan}}\ \emph {et~al.}(1992)\citenamefont
  {{Narayan}}, \citenamefont {{Paczynski}},\ and\ \citenamefont
  {{Piran}}}]{1992ApJ...395L..83N}%
  \BibitemOpen
  \bibfield  {author} {\bibinfo {author} {\bibfnamefont {R.}~\bibnamefont
  {{Narayan}}}, \bibinfo {author} {\bibfnamefont {B.}~\bibnamefont
  {{Paczynski}}}, \ and\ \bibinfo {author} {\bibfnamefont {T.}~\bibnamefont
  {{Piran}}},\ }\href {\doibase 10.1086/186493} {\bibfield  {journal} {\bibinfo
   {journal} {Astrophys. J.}\ }\textbf {\bibinfo {volume} {395}},\ \bibinfo
  {pages} {L83} (\bibinfo {year} {1992})},\ \Eprint
  {http://arxiv.org/abs/astro-ph/9204001} {astro-ph/9204001} \BibitemShut
  {NoStop}%
\bibitem [{\citenamefont {Ferrari}\ \emph {et~al.}(2010)\citenamefont
  {Ferrari}, \citenamefont {Gualtieri},\ and\ \citenamefont
  {Pannarale}}]{PhysRevD.81.064026}%
  \BibitemOpen
  \bibfield  {author} {\bibinfo {author} {\bibfnamefont {V.}~\bibnamefont
  {Ferrari}}, \bibinfo {author} {\bibfnamefont {L.}~\bibnamefont {Gualtieri}},
  \ and\ \bibinfo {author} {\bibfnamefont {F.}~\bibnamefont {Pannarale}},\
  }\href {\doibase 10.1103/PhysRevD.81.064026} {\bibfield  {journal} {\bibinfo
  {journal} {Phys. Rev. D}\ }\textbf {\bibinfo {volume} {81}},\ \bibinfo
  {pages} {064026} (\bibinfo {year} {2010})}\BibitemShut {NoStop}%
\bibitem [{\citenamefont {Pannarale}\ \emph {et~al.}(2011)\citenamefont
  {Pannarale}, \citenamefont {Rezzolla}, \citenamefont {Ohme},\ and\
  \citenamefont {Read}}]{Pannarale:2011pk}%
  \BibitemOpen
  \bibfield  {author} {\bibinfo {author} {\bibfnamefont {F.}~\bibnamefont
  {Pannarale}}, \bibinfo {author} {\bibfnamefont {L.}~\bibnamefont {Rezzolla}},
  \bibinfo {author} {\bibfnamefont {F.}~\bibnamefont {Ohme}}, \ and\ \bibinfo
  {author} {\bibfnamefont {J.~S.}\ \bibnamefont {Read}},\ }\href {\doibase
  10.1103/PhysRevD.84.104017} {\bibfield  {journal} {\bibinfo  {journal} {Phys.
  Rev.}\ }\textbf {\bibinfo {volume} {D84}},\ \bibinfo {pages} {104017}
  (\bibinfo {year} {2011})},\ \Eprint {http://arxiv.org/abs/1103.3526}
  {arXiv:1103.3526 [astro-ph.HE]} \BibitemShut {NoStop}%
\bibitem [{\citenamefont {Etienne}\ \emph {et~al.}(2008)\citenamefont
  {Etienne}, \citenamefont {Faber}, \citenamefont {Liu}, \citenamefont
  {Shapiro}, \citenamefont {Taniguchi},\ and\ \citenamefont
  {Baumgarte}}]{PhysRevD.77.084002}%
  \BibitemOpen
  \bibfield  {author} {\bibinfo {author} {\bibfnamefont {Z.~B.}\ \bibnamefont
  {Etienne}}, \bibinfo {author} {\bibfnamefont {J.~A.}\ \bibnamefont {Faber}},
  \bibinfo {author} {\bibfnamefont {Y.~T.}\ \bibnamefont {Liu}}, \bibinfo
  {author} {\bibfnamefont {S.~L.}\ \bibnamefont {Shapiro}}, \bibinfo {author}
  {\bibfnamefont {K.}~\bibnamefont {Taniguchi}}, \ and\ \bibinfo {author}
  {\bibfnamefont {T.~W.}\ \bibnamefont {Baumgarte}},\ }\href {\doibase
  10.1103/PhysRevD.77.084002} {\bibfield  {journal} {\bibinfo  {journal} {Phys.
  Rev. D}\ }\textbf {\bibinfo {volume} {77}},\ \bibinfo {pages} {084002}
  (\bibinfo {year} {2008})}\BibitemShut {NoStop}%
\bibitem [{\citenamefont {Etienne}\ \emph {et~al.}(2009)\citenamefont
  {Etienne}, \citenamefont {Liu}, \citenamefont {Shapiro},\ and\ \citenamefont
  {Baumgarte}}]{PhysRevD.79.044024}%
  \BibitemOpen
  \bibfield  {author} {\bibinfo {author} {\bibfnamefont {Z.~B.}\ \bibnamefont
  {Etienne}}, \bibinfo {author} {\bibfnamefont {Y.~T.}\ \bibnamefont {Liu}},
  \bibinfo {author} {\bibfnamefont {S.~L.}\ \bibnamefont {Shapiro}}, \ and\
  \bibinfo {author} {\bibfnamefont {T.~W.}\ \bibnamefont {Baumgarte}},\ }\href
  {\doibase 10.1103/PhysRevD.79.044024} {\bibfield  {journal} {\bibinfo
  {journal} {Phys. Rev. D}\ }\textbf {\bibinfo {volume} {79}},\ \bibinfo
  {pages} {044024} (\bibinfo {year} {2009})}\BibitemShut {NoStop}%
\bibitem [{\citenamefont {Duez}\ \emph {et~al.}(2008)\citenamefont {Duez},
  \citenamefont {Foucart}, \citenamefont {Kidder}, \citenamefont {Pfeiffer},
  \citenamefont {Scheel},\ and\ \citenamefont
  {Teukolsky}}]{PhysRevD.78.104015}%
  \BibitemOpen
  \bibfield  {author} {\bibinfo {author} {\bibfnamefont {M.~D.}\ \bibnamefont
  {Duez}}, \bibinfo {author} {\bibfnamefont {F.}~\bibnamefont {Foucart}},
  \bibinfo {author} {\bibfnamefont {L.~E.}\ \bibnamefont {Kidder}}, \bibinfo
  {author} {\bibfnamefont {H.~P.}\ \bibnamefont {Pfeiffer}}, \bibinfo {author}
  {\bibfnamefont {M.~A.}\ \bibnamefont {Scheel}}, \ and\ \bibinfo {author}
  {\bibfnamefont {S.~A.}\ \bibnamefont {Teukolsky}},\ }\href {\doibase
  10.1103/PhysRevD.78.104015} {\bibfield  {journal} {\bibinfo  {journal} {Phys.
  Rev. D}\ }\textbf {\bibinfo {volume} {78}},\ \bibinfo {pages} {104015}
  (\bibinfo {year} {2008})}\BibitemShut {NoStop}%
\bibitem [{\citenamefont {Shibata}\ and\ \citenamefont
  {Ury\ifmmode~\bar{u}\else \={u}\fi{}}(2006)}]{PhysRevD.74.121503}%
  \BibitemOpen
  \bibfield  {author} {\bibinfo {author} {\bibfnamefont {M.}~\bibnamefont
  {Shibata}}\ and\ \bibinfo {author} {\bibfnamefont {K.}~\bibnamefont
  {Ury\ifmmode~\bar{u}\else \={u}\fi{}}},\ }\href {\doibase
  10.1103/PhysRevD.74.121503} {\bibfield  {journal} {\bibinfo  {journal} {Phys.
  Rev. D}\ }\textbf {\bibinfo {volume} {74}},\ \bibinfo {pages} {121503}
  (\bibinfo {year} {2006})}\BibitemShut {NoStop}%
\bibitem [{\citenamefont {L\"offler}\ \emph {et~al.}(2006)\citenamefont
  {L\"offler}, \citenamefont {Rezzolla},\ and\ \citenamefont
  {Ansorg}}]{PhysRevD.74.104018}%
  \BibitemOpen
  \bibfield  {author} {\bibinfo {author} {\bibfnamefont {F.}~\bibnamefont
  {L\"offler}}, \bibinfo {author} {\bibfnamefont {L.}~\bibnamefont {Rezzolla}},
  \ and\ \bibinfo {author} {\bibfnamefont {M.}~\bibnamefont {Ansorg}},\ }\href
  {\doibase 10.1103/PhysRevD.74.104018} {\bibfield  {journal} {\bibinfo
  {journal} {Phys. Rev. D}\ }\textbf {\bibinfo {volume} {74}},\ \bibinfo
  {pages} {104018} (\bibinfo {year} {2006})}\BibitemShut {NoStop}%
\bibitem [{\citenamefont {Apostolatos}\ \emph {et~al.}(1994)\citenamefont
  {Apostolatos}, \citenamefont {Cutler}, \citenamefont {Sussman},\ and\
  \citenamefont {Thorne}}]{Apostolatos:1994mx}%
  \BibitemOpen
  \bibfield  {author} {\bibinfo {author} {\bibfnamefont {T.~A.}\ \bibnamefont
  {Apostolatos}}, \bibinfo {author} {\bibfnamefont {C.}~\bibnamefont {Cutler}},
  \bibinfo {author} {\bibfnamefont {G.~J.}\ \bibnamefont {Sussman}}, \ and\
  \bibinfo {author} {\bibfnamefont {K.~S.}\ \bibnamefont {Thorne}},\ }\href
  {\doibase 10.1103/PhysRevD.49.6274} {\bibfield  {journal} {\bibinfo
  {journal} {Phys. Rev.}\ }\textbf {\bibinfo {volume} {D49}},\ \bibinfo {pages}
  {6274} (\bibinfo {year} {1994})}\BibitemShut {NoStop}%
\bibitem [{\citenamefont {Apostolatos}(1995)}]{Apostolatos:1995pj}%
  \BibitemOpen
  \bibfield  {author} {\bibinfo {author} {\bibfnamefont {T.}~\bibnamefont
  {Apostolatos}},\ }\href {\doibase 10.1103/PhysRevD.52.605} {\bibfield
  {journal} {\bibinfo  {journal} {Phys. Rev.}\ }\textbf {\bibinfo {volume}
  {D52}},\ \bibinfo {pages} {605} (\bibinfo {year} {1995})}\BibitemShut
  {NoStop}%
\bibitem [{\citenamefont {Apostolatos}(1996)}]{Apostolatos:1996rf}%
  \BibitemOpen
  \bibfield  {author} {\bibinfo {author} {\bibfnamefont {T.~A.}\ \bibnamefont
  {Apostolatos}},\ }\href {\doibase 10.1103/PhysRevD.54.2421} {\bibfield
  {journal} {\bibinfo  {journal} {Phys. Rev.}\ }\textbf {\bibinfo {volume}
  {D54}},\ \bibinfo {pages} {2421} (\bibinfo {year} {1996})}\BibitemShut
  {NoStop}%
\bibitem [{\citenamefont {Buonanno}\ \emph {et~al.}(2003)\citenamefont
  {Buonanno}, \citenamefont {Chen},\ and\ \citenamefont
  {Vallisneri}}]{Buonanno:2002fy}%
  \BibitemOpen
  \bibfield  {author} {\bibinfo {author} {\bibfnamefont {A.}~\bibnamefont
  {Buonanno}}, \bibinfo {author} {\bibfnamefont {Y.-b.}\ \bibnamefont {Chen}},
  \ and\ \bibinfo {author} {\bibfnamefont {M.}~\bibnamefont {Vallisneri}},\
  }\href {\doibase 10.1103/PhysRevD.67.104025, 10.1103/PhysRevD.74.029904}
  {\bibfield  {journal} {\bibinfo  {journal} {Phys. Rev.}\ }\textbf {\bibinfo
  {volume} {D67}},\ \bibinfo {pages} {104025} (\bibinfo {year} {2003})},\
  \Eprint {http://arxiv.org/abs/gr-qc/0211087} {arXiv:gr-qc/0211087 [gr-qc]}
  \BibitemShut {NoStop}%
\bibitem [{\citenamefont {Grandclement}\ \emph {et~al.}(2003)\citenamefont
  {Grandclement}, \citenamefont {Kalogera},\ and\ \citenamefont
  {Vecchio}}]{Grandclement:2002dv}%
  \BibitemOpen
  \bibfield  {author} {\bibinfo {author} {\bibfnamefont {P.}~\bibnamefont
  {Grandclement}}, \bibinfo {author} {\bibfnamefont {V.}~\bibnamefont
  {Kalogera}}, \ and\ \bibinfo {author} {\bibfnamefont {A.}~\bibnamefont
  {Vecchio}},\ }\href {\doibase 10.1103/PhysRevD.67.042003} {\bibfield
  {journal} {\bibinfo  {journal} {Phys. Rev.}\ }\textbf {\bibinfo {volume}
  {D67}},\ \bibinfo {pages} {042003} (\bibinfo {year} {2003})},\ \Eprint
  {http://arxiv.org/abs/gr-qc/0207062} {arXiv:gr-qc/0207062 [gr-qc]}
  \BibitemShut {NoStop}%
\bibitem [{\citenamefont {Allen}(2005)}]{Allen:2004gu}%
  \BibitemOpen
  \bibfield  {author} {\bibinfo {author} {\bibfnamefont {B.}~\bibnamefont
  {Allen}},\ }\href {\doibase 10.1103/PhysRevD.71.062001} {\bibfield  {journal}
  {\bibinfo  {journal} {Phys. Rev.}\ }\textbf {\bibinfo {volume} {D71}},\
  \bibinfo {pages} {062001} (\bibinfo {year} {2005})},\ \Eprint
  {http://arxiv.org/abs/gr-qc/0405045} {arXiv:gr-qc/0405045 [gr-qc]}
  \BibitemShut {NoStop}%
\bibitem [{\citenamefont {Babak}\ \emph {et~al.}(2013)\citenamefont {Babak},
  \citenamefont {Biswas}, \citenamefont {Brady}, \citenamefont {Brown},
  \citenamefont {Cannon} \emph {et~al.}}]{Babak:2012zx}%
  \BibitemOpen
  \bibfield  {author} {\bibinfo {author} {\bibfnamefont {S.}~\bibnamefont
  {Babak}}, \bibinfo {author} {\bibfnamefont {R.}~\bibnamefont {Biswas}},
  \bibinfo {author} {\bibfnamefont {P.}~\bibnamefont {Brady}}, \bibinfo
  {author} {\bibfnamefont {D.}~\bibnamefont {Brown}}, \bibinfo {author}
  {\bibfnamefont {K.}~\bibnamefont {Cannon}},  \emph {et~al.},\ }\href
  {\doibase 10.1103/PhysRevD.87.024033} {\bibfield  {journal} {\bibinfo
  {journal} {Phys. Rev.}\ }\textbf {\bibinfo {volume} {D87}},\ \bibinfo {pages}
  {024033} (\bibinfo {year} {2013})},\ \Eprint {http://arxiv.org/abs/1208.3491}
  {arXiv:1208.3491 [gr-qc]} \BibitemShut {NoStop}%
\bibitem [{\citenamefont {Abbott}\ \emph
  {et~al.}(2009{\natexlab{a}})\citenamefont {Abbott} \emph
  {et~al.}}]{Abbott:2009tt}%
  \BibitemOpen
  \bibfield  {author} {\bibinfo {author} {\bibfnamefont {B.}~\bibnamefont
  {Abbott}} \emph {et~al.} (\bibinfo {collaboration} {LIGO Scientific
  Collaboration}),\ }\href {\doibase 10.1103/PhysRevD.79.122001} {\bibfield
  {journal} {\bibinfo  {journal} {Phys. Rev.}\ }\textbf {\bibinfo {volume}
  {D79}},\ \bibinfo {pages} {122001} (\bibinfo {year} {2009}{\natexlab{a}})},\
  \Eprint {http://arxiv.org/abs/0901.0302} {arXiv:0901.0302 [gr-qc]}
  \BibitemShut {NoStop}%
\bibitem [{\citenamefont {Abadie}\ \emph
  {et~al.}(2012{\natexlab{a}})\citenamefont {Abadie} \emph
  {et~al.}}]{Colaboration:2011np}%
  \BibitemOpen
  \bibfield  {author} {\bibinfo {author} {\bibfnamefont {J.}~\bibnamefont
  {Abadie}} \emph {et~al.} (\bibinfo {collaboration} {LIGO Collaboration, Virgo
  Collaboration}),\ }\href {\doibase 10.1103/PhysRevD.85.082002} {\bibfield
  {journal} {\bibinfo  {journal} {Phys. Rev.}\ }\textbf {\bibinfo {volume}
  {D85}},\ \bibinfo {pages} {082002} (\bibinfo {year} {2012}{\natexlab{a}})},\
  \Eprint {http://arxiv.org/abs/1111.7314} {arXiv:1111.7314 [gr-qc]}
  \BibitemShut {NoStop}%
\bibitem [{\citenamefont {Abadie}\ \emph {et~al.}(2011)\citenamefont {Abadie}
  \emph {et~al.}}]{Abadie:2011kd}%
  \BibitemOpen
  \bibfield  {author} {\bibinfo {author} {\bibfnamefont {J.}~\bibnamefont
  {Abadie}} \emph {et~al.} (\bibinfo {collaboration} {LIGO Scientific
  Collaboration, Virgo Collaboration}),\ }\href {\doibase
  10.1103/PhysRevD.86.069903, 10.1103/PhysRevD.85.089904,
  10.1103/PhysRevD.83.122005} {\bibfield  {journal} {\bibinfo  {journal} {Phys.
  Rev.}\ }\textbf {\bibinfo {volume} {D83}},\ \bibinfo {pages} {122005}
  (\bibinfo {year} {2011})},\ \Eprint {http://arxiv.org/abs/1102.3781}
  {arXiv:1102.3781 [gr-qc]} \BibitemShut {NoStop}%
\bibitem [{\citenamefont {Abadie}\ \emph
  {et~al.}(2010{\natexlab{b}})\citenamefont {Abadie} \emph
  {et~al.}}]{Abadie:2010yb}%
  \BibitemOpen
  \bibfield  {author} {\bibinfo {author} {\bibfnamefont {J.}~\bibnamefont
  {Abadie}} \emph {et~al.} (\bibinfo {collaboration} {LIGO Scientific
  Collaboration, Virgo Collaboration}),\ }\href {\doibase
  10.1103/PhysRevD.85.089903, 10.1103/PhysRevD.82.102001} {\bibfield  {journal}
  {\bibinfo  {journal} {Phys. Rev.}\ }\textbf {\bibinfo {volume} {D82}},\
  \bibinfo {pages} {102001} (\bibinfo {year} {2010}{\natexlab{b}})},\ \Eprint
  {http://arxiv.org/abs/1005.4655} {arXiv:1005.4655 [gr-qc]} \BibitemShut
  {NoStop}%
\bibitem [{\citenamefont {Abbott}\ \emph
  {et~al.}(2009{\natexlab{b}})\citenamefont {Abbott} \emph
  {et~al.}}]{Abbott:2009qj}%
  \BibitemOpen
  \bibfield  {author} {\bibinfo {author} {\bibfnamefont {B.}~\bibnamefont
  {Abbott}} \emph {et~al.} (\bibinfo {collaboration} {LIGO Scientific
  Collaboration}),\ }\href {\doibase 10.1103/PhysRevD.80.047101} {\bibfield
  {journal} {\bibinfo  {journal} {Phys. Rev.}\ }\textbf {\bibinfo {volume}
  {D80}},\ \bibinfo {pages} {047101} (\bibinfo {year} {2009}{\natexlab{b}})},\
  \Eprint {http://arxiv.org/abs/0905.3710} {arXiv:0905.3710 [gr-qc]}
  \BibitemShut {NoStop}%
\bibitem [{\citenamefont {Abbott}\ \emph {et~al.}(2006)\citenamefont {Abbott}
  \emph {et~al.}}]{PRD.73.102002}%
  \BibitemOpen
  \bibfield  {author} {\bibinfo {author} {\bibfnamefont {B.}~\bibnamefont
  {Abbott}} \emph {et~al.} (\bibinfo {collaboration} {LIGO Scientific
  Collaboration, http://www.ligo.org and TAMA Collaboration}),\ }\href
  {\doibase 10.1103/PhysRevD.73.102002} {\bibfield  {journal} {\bibinfo
  {journal} {Phys. Rev. D}\ }\textbf {\bibinfo {volume} {73}},\ \bibinfo
  {pages} {102002} (\bibinfo {year} {2006})}\BibitemShut {NoStop}%
\bibitem [{\citenamefont {Abbott}\ \emph {et~al.}(2004)\citenamefont {Abbott}
  \emph {et~al.}}]{PRD.69.122001}%
  \BibitemOpen
  \bibfield  {author} {\bibinfo {author} {\bibfnamefont {B.}~\bibnamefont
  {Abbott}} \emph {et~al.} (\bibinfo {collaboration} {(LIGO Scientific
  Collaboration)}),\ }\href {\doibase 10.1103/PhysRevD.69.122001} {\bibfield
  {journal} {\bibinfo  {journal} {Phys. Rev. D}\ }\textbf {\bibinfo {volume}
  {69}},\ \bibinfo {pages} {122001} (\bibinfo {year} {2004})}\BibitemShut
  {NoStop}%
\bibitem [{\citenamefont {Abbott}\ \emph
  {et~al.}(2005{\natexlab{a}})\citenamefont {Abbott} \emph
  {et~al.}}]{PRD.72.082001}%
  \BibitemOpen
  \bibfield  {author} {\bibinfo {author} {\bibfnamefont {B.}~\bibnamefont
  {Abbott}} \emph {et~al.} (\bibinfo {collaboration} {LIGO Scientific
  Collaboration}),\ }\href {\doibase 10.1103/PhysRevD.72.082001} {\bibfield
  {journal} {\bibinfo  {journal} {Phys. Rev. D}\ }\textbf {\bibinfo {volume}
  {72}},\ \bibinfo {pages} {082001} (\bibinfo {year}
  {2005}{\natexlab{a}})}\BibitemShut {NoStop}%
\bibitem [{\citenamefont {Abbott}\ \emph
  {et~al.}(2005{\natexlab{b}})\citenamefont {Abbott} \emph
  {et~al.}}]{PRD.72.082002}%
  \BibitemOpen
  \bibfield  {author} {\bibinfo {author} {\bibfnamefont {B.}~\bibnamefont
  {Abbott}} \emph {et~al.} (\bibinfo {collaboration} {LIGO Scientific
  Collaboration}),\ }\href {\doibase 10.1103/PhysRevD.72.082002} {\bibfield
  {journal} {\bibinfo  {journal} {Phys. Rev. D}\ }\textbf {\bibinfo {volume}
  {72}},\ \bibinfo {pages} {082002} (\bibinfo {year}
  {2005}{\natexlab{b}})}\BibitemShut {NoStop}%
\bibitem [{\citenamefont {Abadie}\ \emph
  {et~al.}(2010{\natexlab{c}})\citenamefont {Abadie} \emph
  {et~al.}}]{0004-637X-715-2-1453}%
  \BibitemOpen
  \bibfield  {author} {\bibinfo {author} {\bibfnamefont {J.}~\bibnamefont
  {Abadie}} \emph {et~al.} (\bibinfo {collaboration} {LIGO Scientific
  Collaboration, Virgo Collaboration}),\ }\href
  {http://stacks.iop.org/0004-637X/715/i=2/a=1453} {\bibfield  {journal}
  {\bibinfo  {journal} {The Astrophysical Journal}\ }\textbf {\bibinfo {volume}
  {715}},\ \bibinfo {pages} {1453} (\bibinfo {year}
  {2010}{\natexlab{c}})}\BibitemShut {NoStop}%
\bibitem [{\citenamefont {Aasi}\ \emph
  {et~al.}(2013{\natexlab{b}})\citenamefont {Aasi} \emph
  {et~al.}}]{Aasi:2013wya}%
  \BibitemOpen
  \bibfield  {author} {\bibinfo {author} {\bibfnamefont {J.}~\bibnamefont
  {Aasi}} \emph {et~al.} (\bibinfo {collaboration} {LIGO Scientific
  Collaboration, Virgo Collaboration}),\ }\href@noop {} {\  (\bibinfo {year}
  {2013}{\natexlab{b}})},\ \Eprint {http://arxiv.org/abs/1304.0670}
  {arXiv:1304.0670 [gr-qc]} \BibitemShut {NoStop}%
\bibitem [{\citenamefont {Thorne}(1974)}]{Thorne:1974ve}%
  \BibitemOpen
  \bibfield  {author} {\bibinfo {author} {\bibfnamefont {K.~S.}\ \bibnamefont
  {Thorne}},\ }\href {\doibase 10.1086/152991} {\bibfield  {journal} {\bibinfo
  {journal} {Astrophys.J.}\ }\textbf {\bibinfo {volume} {191}},\ \bibinfo
  {pages} {507} (\bibinfo {year} {1974})}\BibitemShut {NoStop}%
\bibitem [{\citenamefont {McClintock}\ \emph {et~al.}(2013)\citenamefont
  {McClintock}, \citenamefont {Narayan},\ and\ \citenamefont
  {Steiner}}]{McClintock:2013vwa}%
  \BibitemOpen
  \bibfield  {author} {\bibinfo {author} {\bibfnamefont {J.~E.}\ \bibnamefont
  {McClintock}}, \bibinfo {author} {\bibfnamefont {R.}~\bibnamefont {Narayan}},
  \ and\ \bibinfo {author} {\bibfnamefont {J.~F.}\ \bibnamefont {Steiner}},\
  }\href@noop {} {\  (\bibinfo {year} {2013})},\ \Eprint
  {http://arxiv.org/abs/1303.1583} {arXiv:1303.1583 [astro-ph.HE]} \BibitemShut
  {NoStop}%
\bibitem [{\citenamefont {Dominik}\ \emph {et~al.}(2012)\citenamefont
  {Dominik}, \citenamefont {Belczynski}, \citenamefont {Fryer}, \citenamefont
  {Holz}, \citenamefont {Berti} \emph {et~al.}}]{Dominik:2012kk}%
  \BibitemOpen
  \bibfield  {author} {\bibinfo {author} {\bibfnamefont {M.}~\bibnamefont
  {Dominik}}, \bibinfo {author} {\bibfnamefont {K.}~\bibnamefont {Belczynski}},
  \bibinfo {author} {\bibfnamefont {C.}~\bibnamefont {Fryer}}, \bibinfo
  {author} {\bibfnamefont {D.}~\bibnamefont {Holz}}, \bibinfo {author}
  {\bibfnamefont {E.}~\bibnamefont {Berti}},  \emph {et~al.},\ }\href {\doibase
  10.1088/0004-637X/759/1/52} {\bibfield  {journal} {\bibinfo  {journal}
  {Astrophys.J.}\ }\textbf {\bibinfo {volume} {759}},\ \bibinfo {pages} {52}
  (\bibinfo {year} {2012})},\ \Eprint {http://arxiv.org/abs/1202.4901}
  {arXiv:1202.4901 [astro-ph.HE]} \BibitemShut {NoStop}%
\bibitem [{\citenamefont {Lo}\ and\ \citenamefont {Lin}(2011)}]{Lo:2010bj}%
  \BibitemOpen
  \bibfield  {author} {\bibinfo {author} {\bibfnamefont {K.-W.}\ \bibnamefont
  {Lo}}\ and\ \bibinfo {author} {\bibfnamefont {L.-M.}\ \bibnamefont {Lin}},\
  }\href {\doibase 10.1088/0004-637X/728/1/12} {\bibfield  {journal} {\bibinfo
  {journal} {Astrophys.J.}\ }\textbf {\bibinfo {volume} {728}},\ \bibinfo
  {pages} {12} (\bibinfo {year} {2011})},\ \Eprint
  {http://arxiv.org/abs/1011.3563} {arXiv:1011.3563 [astro-ph.HE]} \BibitemShut
  {NoStop}%
\bibitem [{\citenamefont {Wagoner}(1984)}]{Wagoner:1984pv}%
  \BibitemOpen
  \bibfield  {author} {\bibinfo {author} {\bibfnamefont {R.}~\bibnamefont
  {Wagoner}},\ }\href {\doibase 10.1086/161798} {\bibfield  {journal} {\bibinfo
   {journal} {Astrophys.J.}\ }\textbf {\bibinfo {volume} {278}},\ \bibinfo
  {pages} {345} (\bibinfo {year} {1984})}\BibitemShut {NoStop}%
\bibitem [{\citenamefont {Bildsten}(1998)}]{Bildsten:1998ey}%
  \BibitemOpen
  \bibfield  {author} {\bibinfo {author} {\bibfnamefont {L.}~\bibnamefont
  {Bildsten}},\ }\href {\doibase 10.1086/311440} {\bibfield  {journal}
  {\bibinfo  {journal} {Astrophys.J.}\ }\textbf {\bibinfo {volume} {501}},\
  \bibinfo {pages} {L89} (\bibinfo {year} {1998})},\ \Eprint
  {http://arxiv.org/abs/astro-ph/9804325} {arXiv:astro-ph/9804325 [astro-ph]}
  \BibitemShut {NoStop}%
\bibitem [{\citenamefont {Chakrabarty}\ \emph {et~al.}(2003)\citenamefont
  {Chakrabarty}, \citenamefont {Morgan}, \citenamefont {Muno}, \citenamefont
  {Galloway}, \citenamefont {Wijnands} \emph {et~al.}}]{Chakrabarty:2003kt}%
  \BibitemOpen
  \bibfield  {author} {\bibinfo {author} {\bibfnamefont {D.}~\bibnamefont
  {Chakrabarty}}, \bibinfo {author} {\bibfnamefont {E.~H.}\ \bibnamefont
  {Morgan}}, \bibinfo {author} {\bibfnamefont {M.~P.}\ \bibnamefont {Muno}},
  \bibinfo {author} {\bibfnamefont {D.~K.}\ \bibnamefont {Galloway}}, \bibinfo
  {author} {\bibfnamefont {R.}~\bibnamefont {Wijnands}},  \emph {et~al.},\
  }\href {\doibase 10.1038/nature01732} {\bibfield  {journal} {\bibinfo
  {journal} {Nature}\ }\textbf {\bibinfo {volume} {424}},\ \bibinfo {pages}
  {42} (\bibinfo {year} {2003})},\ \Eprint
  {http://arxiv.org/abs/astro-ph/0307029} {arXiv:astro-ph/0307029 [astro-ph]}
  \BibitemShut {NoStop}%
\bibitem [{\citenamefont {Damour}\ \emph {et~al.}(2012)\citenamefont {Damour},
  \citenamefont {Nagar},\ and\ \citenamefont {Villain}}]{Damour:2012yf}%
  \BibitemOpen
  \bibfield  {author} {\bibinfo {author} {\bibfnamefont {T.}~\bibnamefont
  {Damour}}, \bibinfo {author} {\bibfnamefont {A.}~\bibnamefont {Nagar}}, \
  and\ \bibinfo {author} {\bibfnamefont {L.}~\bibnamefont {Villain}},\ }\href
  {\doibase 10.1103/PhysRevD.85.123007} {\bibfield  {journal} {\bibinfo
  {journal} {Phys. Rev.}\ }\textbf {\bibinfo {volume} {D85}},\ \bibinfo {pages}
  {123007} (\bibinfo {year} {2012})},\ \Eprint {http://arxiv.org/abs/1203.4352}
  {arXiv:1203.4352 [gr-qc]} \BibitemShut {NoStop}%
\bibitem [{\citenamefont {Belczynski}\ \emph {et~al.}(2007)\citenamefont
  {Belczynski}, \citenamefont {Taam}, \citenamefont {Rantsiou},\ and\
  \citenamefont {van~der Sluys}}]{Belczynski:2007xg}%
  \BibitemOpen
  \bibfield  {author} {\bibinfo {author} {\bibfnamefont {K.}~\bibnamefont
  {Belczynski}}, \bibinfo {author} {\bibfnamefont {R.~E.}\ \bibnamefont
  {Taam}}, \bibinfo {author} {\bibfnamefont {E.}~\bibnamefont {Rantsiou}}, \
  and\ \bibinfo {author} {\bibfnamefont {M.}~\bibnamefont {van~der Sluys}},\
  }\href@noop {} {\bibfield  {journal} {\bibinfo  {journal} {Astrophys.J.}\ }
  (\bibinfo {year} {2007})},\ \Eprint {http://arxiv.org/abs/astro-ph/0703131}
  {arXiv:astro-ph/0703131 [ASTRO-PH]} \BibitemShut {NoStop}%
\bibitem [{\citenamefont {Kalogera}(2000)}]{Kalogera:1999tq}%
  \BibitemOpen
  \bibfield  {author} {\bibinfo {author} {\bibfnamefont {V.}~\bibnamefont
  {Kalogera}},\ }\href {\doibase 10.1086/309400} {\bibfield  {journal}
  {\bibinfo  {journal} {Astrophys.J.}\ }\textbf {\bibinfo {volume} {541}},\
  \bibinfo {pages} {319} (\bibinfo {year} {2000})},\ \Eprint
  {http://arxiv.org/abs/astro-ph/9911417} {arXiv:astro-ph/9911417 [astro-ph]}
  \BibitemShut {NoStop}%
\bibitem [{\citenamefont {Lundgren}\ and\ \citenamefont
  {O'Shaughnessy}(2013)}]{Lundgren:2013jla}%
  \BibitemOpen
  \bibfield  {author} {\bibinfo {author} {\bibfnamefont {A.}~\bibnamefont
  {Lundgren}}\ and\ \bibinfo {author} {\bibfnamefont {R.}~\bibnamefont
  {O'Shaughnessy}},\ }\href@noop {} {\  (\bibinfo {year} {2013})},\ \Eprint
  {http://arxiv.org/abs/1304.3332} {arXiv:1304.3332 [gr-qc]} \BibitemShut
  {NoStop}%
\bibitem [{\citenamefont {Foucart}\ \emph {et~al.}(2013)\citenamefont
  {Foucart}, \citenamefont {Deaton}, \citenamefont {Duez}, \citenamefont
  {Kidder}, \citenamefont {MacDonald} \emph {et~al.}}]{Foucart:2012vn}%
  \BibitemOpen
  \bibfield  {author} {\bibinfo {author} {\bibfnamefont {F.}~\bibnamefont
  {Foucart}}, \bibinfo {author} {\bibfnamefont {M.~B.}\ \bibnamefont {Deaton}},
  \bibinfo {author} {\bibfnamefont {M.~D.}\ \bibnamefont {Duez}}, \bibinfo
  {author} {\bibfnamefont {L.~E.}\ \bibnamefont {Kidder}}, \bibinfo {author}
  {\bibfnamefont {I.}~\bibnamefont {MacDonald}},  \emph {et~al.},\ }\href
  {\doibase 10.1103/PhysRevD.87.084006} {\bibfield  {journal} {\bibinfo
  {journal} {Phys. Rev.}\ }\textbf {\bibinfo {volume} {D87}},\ \bibinfo {pages}
  {084006} (\bibinfo {year} {2013})},\ \Eprint {http://arxiv.org/abs/1212.4810}
  {arXiv:1212.4810 [gr-qc]} \BibitemShut {NoStop}%
\bibitem [{\citenamefont {Pan}\ \emph {et~al.}(2011)\citenamefont {Pan},
  \citenamefont {Buonanno}, \citenamefont {Boyle}, \citenamefont {Buchman},
  \citenamefont {Kidder} \emph {et~al.}}]{Pan:2011gk}%
  \BibitemOpen
  \bibfield  {author} {\bibinfo {author} {\bibfnamefont {Y.}~\bibnamefont
  {Pan}}, \bibinfo {author} {\bibfnamefont {A.}~\bibnamefont {Buonanno}},
  \bibinfo {author} {\bibfnamefont {M.}~\bibnamefont {Boyle}}, \bibinfo
  {author} {\bibfnamefont {L.~T.}\ \bibnamefont {Buchman}}, \bibinfo {author}
  {\bibfnamefont {L.~E.}\ \bibnamefont {Kidder}},  \emph {et~al.},\ }\href
  {\doibase 10.1103/PhysRevD.84.124052} {\bibfield  {journal} {\bibinfo
  {journal} {Phys. Rev.}\ }\textbf {\bibinfo {volume} {D84}},\ \bibinfo {pages}
  {124052} (\bibinfo {year} {2011})},\ \Eprint {http://arxiv.org/abs/1106.1021}
  {arXiv:1106.1021 [gr-qc]} \BibitemShut {NoStop}%
\bibitem [{\citenamefont {Brown}\ \emph {et~al.}(2013)\citenamefont {Brown},
  \citenamefont {Kumar},\ and\ \citenamefont {Nitz}}]{Brown:2012nn}%
  \BibitemOpen
  \bibfield  {author} {\bibinfo {author} {\bibfnamefont {D.~A.}\ \bibnamefont
  {Brown}}, \bibinfo {author} {\bibfnamefont {P.}~\bibnamefont {Kumar}}, \ and\
  \bibinfo {author} {\bibfnamefont {A.~H.}\ \bibnamefont {Nitz}},\ }\href
  {\doibase 10.1103/PhysRevD.87.082004} {\bibfield  {journal} {\bibinfo
  {journal} {Phys. Rev.}\ }\textbf {\bibinfo {volume} {D87}},\ \bibinfo {pages}
  {082004} (\bibinfo {year} {2013})},\ \Eprint {http://arxiv.org/abs/1211.6184}
  {arXiv:1211.6184 [gr-qc]} \BibitemShut {NoStop}%
\bibitem [{pyc({\natexlab{a}})}]{pycbc}%
  \BibitemOpen
  \href@noop {} {\enquote {\bibinfo {title} {{PyCBC}},}\ }\bibinfo
  {howpublished}
  {\url{https://www.lsc-group.phys.uwm.edu/daswg/projects/pycbc.html}}
  ({\natexlab{a}})\BibitemShut {NoStop}%
\bibitem [{LAL()}]{LAL}%
  \BibitemOpen
  \href@noop {} {\enquote {\bibinfo {title} {{The LIGO Algorithms Library}},}\
  }\bibinfo {howpublished}
  {\url{https://www.lsc-group.phys.uwm.edu/daswg/projects/lalsuite.html}}\BibitemShut
  {NoStop}%
\bibitem [{\citenamefont {Aasi}\ \emph {et~al.}(2014)\citenamefont {Aasi} \emph
  {et~al.}}]{Aasi:2014tra}%
  \BibitemOpen
  \bibfield  {author} {\bibinfo {author} {\bibfnamefont {J.}~\bibnamefont
  {Aasi}} \emph {et~al.} (\bibinfo {collaboration} {The LIGO Scientific
  Collaboration, the Virgo Collaboration, the NINJA-2 Collaboration}),\
  }\href@noop {} {\  (\bibinfo {year} {2014})},\ \Eprint
  {http://arxiv.org/abs/1401.0939} {arXiv:1401.0939 [gr-qc]} \BibitemShut
  {NoStop}%
\bibitem [{\citenamefont {Allen}\ \emph {et~al.}(2012)\citenamefont {Allen},
  \citenamefont {Anderson}, \citenamefont {Brady}, \citenamefont {Brown},\ and\
  \citenamefont {Creighton}}]{Allen:2005fk}%
  \BibitemOpen
  \bibfield  {author} {\bibinfo {author} {\bibfnamefont {B.}~\bibnamefont
  {Allen}}, \bibinfo {author} {\bibfnamefont {W.~G.}\ \bibnamefont {Anderson}},
  \bibinfo {author} {\bibfnamefont {P.~R.}\ \bibnamefont {Brady}}, \bibinfo
  {author} {\bibfnamefont {D.~A.}\ \bibnamefont {Brown}}, \ and\ \bibinfo
  {author} {\bibfnamefont {J.~D.}\ \bibnamefont {Creighton}},\ }\href {\doibase
  10.1103/PhysRevD.85.122006} {\bibfield  {journal} {\bibinfo  {journal} {Phys.
  Rev.}\ }\textbf {\bibinfo {volume} {D85}},\ \bibinfo {pages} {122006}
  (\bibinfo {year} {2012})},\ \Eprint {http://arxiv.org/abs/gr-qc/0509116}
  {arXiv:gr-qc/0509116 [gr-qc]} \BibitemShut {NoStop}%
\bibitem [{\citenamefont {Percival}\ and\ \citenamefont
  {Walden}(1993)}]{PercivalWalden}%
  \BibitemOpen
  \bibfield  {author} {\bibinfo {author} {\bibfnamefont {D.}~\bibnamefont
  {Percival}}\ and\ \bibinfo {author} {\bibfnamefont {A.}~\bibnamefont
  {Walden}},\ }\href@noop {} {\emph {\bibinfo {title} {Spectral Analysis for
  Physical Applications: Multitaper and Conventional Univariate Techniques}}}\
  (\bibinfo  {publisher} {Cambridge University Press},\ \bibinfo {address}
  {Cambridge, UK},\ \bibinfo {year} {1993})\BibitemShut {NoStop}%
\bibitem [{\citenamefont {Abadie}\ \emph
  {et~al.}(2012{\natexlab{b}})\citenamefont {Abadie} \emph
  {et~al.}}]{LIGO:2012aa}%
  \BibitemOpen
  \bibfield  {author} {\bibinfo {author} {\bibfnamefont {J.}~\bibnamefont
  {Abadie}} \emph {et~al.} (\bibinfo {collaboration} {Virgo Collaboration, LIGO
  Scientific Collaboration}),\ }\href@noop {} {\  (\bibinfo {year}
  {2012}{\natexlab{b}})},\ \Eprint {http://arxiv.org/abs/1203.2674}
  {arXiv:1203.2674 [gr-qc]} \BibitemShut {NoStop}%
\bibitem [{\citenamefont {Cutler}\ and\ \citenamefont
  {Flanagan}(1994)}]{PhysRevD.49.2658}%
  \BibitemOpen
  \bibfield  {author} {\bibinfo {author} {\bibfnamefont {C.}~\bibnamefont
  {Cutler}}\ and\ \bibinfo {author} {\bibfnamefont {E.~E.}\ \bibnamefont
  {Flanagan}},\ }\href {\doibase 10.1103/PhysRevD.49.2658} {\bibfield
  {journal} {\bibinfo  {journal} {Phys. Rev. D}\ }\textbf {\bibinfo {volume}
  {49}},\ \bibinfo {pages} {2658} (\bibinfo {year} {1994})}\BibitemShut
  {NoStop}%
\bibitem [{pyt()}]{python}%
  \BibitemOpen
  \href@noop {} {\enquote {\bibinfo {title} {{The Python Programming
  Language}},}\ }\bibinfo {howpublished}
  {\url{http://www.python.org}}\BibitemShut {NoStop}%
\bibitem [{swi()}]{swig}%
  \BibitemOpen
  \href@noop {} {\enquote {\bibinfo {title} {{SWIG}},}\ }\bibinfo
  {howpublished} {\url{http://www.swig.org}}\BibitemShut {NoStop}%
\bibitem [{cud()}]{cuda}%
  \BibitemOpen
  \href@noop {} {\enquote {\bibinfo {title} {{CUDA}},}\ }\bibinfo
  {howpublished}
  {\url{http://www.nvidia.com/object/cuda\_home\_new.html}}\BibitemShut
  {NoStop}%
\bibitem [{ope()}]{opencl}%
  \BibitemOpen
  \href@noop {} {\enquote {\bibinfo {title} {{OpenCL}},}\ }\bibinfo
  {howpublished} {\url{https://www.khronos.org/opencl}}\BibitemShut {NoStop}%
\bibitem [{pyc({\natexlab{b}})}]{pycuda}%
  \BibitemOpen
  \href@noop {} {\enquote {\bibinfo {title} {{PyCUDA}},}\ }\bibinfo
  {howpublished} {\url{http://mathema.tician.de/software/pycuda}}
  ({\natexlab{b}})\BibitemShut {NoStop}%
\bibitem [{pyo()}]{pyopencl}%
  \BibitemOpen
  \href@noop {} {\enquote {\bibinfo {title} {{PyOpenCL}},}\ }\bibinfo
  {howpublished} {\url{http://mathema.tician.de/software/pyopencl}}\BibitemShut
  {NoStop}%
\bibitem [{cuf()}]{cufft}%
  \BibitemOpen
  \href@noop {} {\enquote {\bibinfo {title} {{cuFFT}},}\ }\bibinfo
  {howpublished} {\url{https://developer.nvidia.com/cuFFT}}\BibitemShut
  {NoStop}%
\bibitem [{\citenamefont {Frigo}\ and\ \citenamefont {Johnson}(2005)}]{fftw}%
  \BibitemOpen
  \bibfield  {author} {\bibinfo {author} {\bibfnamefont {M.}~\bibnamefont
  {Frigo}}\ and\ \bibinfo {author} {\bibfnamefont {S.~G.}\ \bibnamefont
  {Johnson}},\ }\href@noop {} {\bibfield  {journal} {\bibinfo  {journal}
  {Proceedings of the IEEE}\ }\textbf {\bibinfo {volume} {93}},\ \bibinfo
  {pages} {216} (\bibinfo {year} {2005})},\ \bibinfo {note} {special issue on
  ``Program Generation, Optimization, and Platform Adaptation''}\BibitemShut
  {NoStop}%
\bibitem [{mkl()}]{mkl}%
  \BibitemOpen
  \href@noop {} {\enquote {\bibinfo {title} {{MKL}},}\ }\bibinfo {howpublished}
  {\url{http://software.intel.com/en-us/intel-mkl}}\BibitemShut {NoStop}%
\bibitem [{atl()}]{atlas}%
  \BibitemOpen
  \href@noop {} {\enquote {\bibinfo {title} {{The Atlas Computational
  Cluster}},}\ }\bibinfo {howpublished}
  {\url{https://wiki.atlas.aei.uni-hannover.de/foswiki/bin/view/ATLAS/WebHome}}\BibitemShut
  {NoStop}%
\bibitem [{nvi()}]{nvidiasmi}%
  \BibitemOpen
  \href@noop {} {\enquote {\bibinfo {title} {{NVIDIA System Management
  Interface}},}\ }\bibinfo {howpublished}
  {\url{https://developer.nvidia.com/nvidia-system-management-interface}}\BibitemShut
  {NoStop}%
\bibitem [{\citenamefont {Harry}\ \emph {et~al.}(2009)\citenamefont {Harry},
  \citenamefont {Allen},\ and\ \citenamefont {Sathyaprakash}}]{Harry:2009ea}%
  \BibitemOpen
  \bibfield  {author} {\bibinfo {author} {\bibfnamefont {I.~W.}\ \bibnamefont
  {Harry}}, \bibinfo {author} {\bibfnamefont {B.}~\bibnamefont {Allen}}, \ and\
  \bibinfo {author} {\bibfnamefont {B.}~\bibnamefont {Sathyaprakash}},\ }\href
  {\doibase 10.1103/PhysRevD.80.104014} {\bibfield  {journal} {\bibinfo
  {journal} {Phys. Rev.}\ }\textbf {\bibinfo {volume} {D80}},\ \bibinfo {pages}
  {104014} (\bibinfo {year} {2009})},\ \Eprint {http://arxiv.org/abs/0908.2090}
  {arXiv:0908.2090 [gr-qc]} \BibitemShut {NoStop}%
\bibitem [{\citenamefont {Babak}(2008)}]{Babak:2008rb}%
  \BibitemOpen
  \bibfield  {author} {\bibinfo {author} {\bibfnamefont {S.}~\bibnamefont
  {Babak}},\ }\href {\doibase 10.1088/0264-9381/25/19/195011} {\bibfield
  {journal} {\bibinfo  {journal} {Class.Quant.Grav.}\ }\textbf {\bibinfo
  {volume} {25}},\ \bibinfo {pages} {195011} (\bibinfo {year} {2008})},\
  \Eprint {http://arxiv.org/abs/0801.4070} {arXiv:0801.4070 [gr-qc]}
  \BibitemShut {NoStop}%
\bibitem [{\citenamefont {Blanchet}\ \emph {et~al.}(1995)\citenamefont
  {Blanchet}, \citenamefont {Damour}, \citenamefont {Iyer}, \citenamefont
  {Will},\ and\ \citenamefont {Wiseman}}]{PhysRevLett.74.3515}%
  \BibitemOpen
  \bibfield  {author} {\bibinfo {author} {\bibfnamefont {L.}~\bibnamefont
  {Blanchet}}, \bibinfo {author} {\bibfnamefont {T.}~\bibnamefont {Damour}},
  \bibinfo {author} {\bibfnamefont {B.~R.}\ \bibnamefont {Iyer}}, \bibinfo
  {author} {\bibfnamefont {C.~M.}\ \bibnamefont {Will}}, \ and\ \bibinfo
  {author} {\bibfnamefont {A.~G.}\ \bibnamefont {Wiseman}},\ }\href {\doibase
  10.1103/PhysRevLett.74.3515} {\bibfield  {journal} {\bibinfo  {journal}
  {Phys. Rev. Lett.}\ }\textbf {\bibinfo {volume} {74}},\ \bibinfo {pages}
  {3515} (\bibinfo {year} {1995})}\BibitemShut {NoStop}%
\bibitem [{\citenamefont {Blanchet}\ \emph {et~al.}(2004)\citenamefont
  {Blanchet}, \citenamefont {Damour}, \citenamefont {Esposito-Far\`ese},\ and\
  \citenamefont {Iyer}}]{PhysRevLett.93.091101}%
  \BibitemOpen
  \bibfield  {author} {\bibinfo {author} {\bibfnamefont {L.}~\bibnamefont
  {Blanchet}}, \bibinfo {author} {\bibfnamefont {T.}~\bibnamefont {Damour}},
  \bibinfo {author} {\bibfnamefont {G.}~\bibnamefont {Esposito-Far\`ese}}, \
  and\ \bibinfo {author} {\bibfnamefont {B.~R.}\ \bibnamefont {Iyer}},\ }\href
  {\doibase 10.1103/PhysRevLett.93.091101} {\bibfield  {journal} {\bibinfo
  {journal} {Phys. Rev. Lett.}\ }\textbf {\bibinfo {volume} {93}},\ \bibinfo
  {pages} {091101} (\bibinfo {year} {2004})}\BibitemShut {NoStop}%
\bibitem [{\citenamefont {Arun}\ \emph {et~al.}(2009)\citenamefont {Arun},
  \citenamefont {Buonanno}, \citenamefont {Faye},\ and\ \citenamefont
  {Ochsner}}]{Arun:2008kb}%
  \BibitemOpen
  \bibfield  {author} {\bibinfo {author} {\bibfnamefont {K.}~\bibnamefont
  {Arun}}, \bibinfo {author} {\bibfnamefont {A.}~\bibnamefont {Buonanno}},
  \bibinfo {author} {\bibfnamefont {G.}~\bibnamefont {Faye}}, \ and\ \bibinfo
  {author} {\bibfnamefont {E.}~\bibnamefont {Ochsner}},\ }\href {\doibase
  10.1103/PhysRevD.79.104023, 10.1103/PhysRevD.84.049901} {\bibfield  {journal}
  {\bibinfo  {journal} {Phys. Rev.}\ }\textbf {\bibinfo {volume} {D79}},\
  \bibinfo {pages} {104023} (\bibinfo {year} {2009})},\ \Eprint
  {http://arxiv.org/abs/0810.5336} {arXiv:0810.5336 [gr-qc]} \BibitemShut
  {NoStop}%
\bibitem [{\citenamefont {Kidder}(1995)}]{PhysRevD.52.821}%
  \BibitemOpen
  \bibfield  {author} {\bibinfo {author} {\bibfnamefont {L.~E.}\ \bibnamefont
  {Kidder}},\ }\href {\doibase 10.1103/PhysRevD.52.821} {\bibfield  {journal}
  {\bibinfo  {journal} {Phys. Rev. D}\ }\textbf {\bibinfo {volume} {52}},\
  \bibinfo {pages} {821} (\bibinfo {year} {1995})}\BibitemShut {NoStop}%
\bibitem [{\citenamefont {Blanchet}\ \emph {et~al.}(2006)\citenamefont
  {Blanchet}, \citenamefont {Buonanno},\ and\ \citenamefont
  {Faye}}]{PhysRevD.74.104034}%
  \BibitemOpen
  \bibfield  {author} {\bibinfo {author} {\bibfnamefont {L.}~\bibnamefont
  {Blanchet}}, \bibinfo {author} {\bibfnamefont {A.}~\bibnamefont {Buonanno}},
  \ and\ \bibinfo {author} {\bibfnamefont {G.}~\bibnamefont {Faye}},\ }\href
  {\doibase 10.1103/PhysRevD.74.104034} {\bibfield  {journal} {\bibinfo
  {journal} {Phys. Rev. D}\ }\textbf {\bibinfo {volume} {74}},\ \bibinfo
  {pages} {104034} (\bibinfo {year} {2006})}\BibitemShut {NoStop}%
\bibitem [{\citenamefont {Baird}\ \emph {et~al.}(2013)\citenamefont {Baird},
  \citenamefont {Fairhurst}, \citenamefont {Hannam},\ and\ \citenamefont
  {Murphy}}]{PhysRevD.87.024035}%
  \BibitemOpen
  \bibfield  {author} {\bibinfo {author} {\bibfnamefont {E.}~\bibnamefont
  {Baird}}, \bibinfo {author} {\bibfnamefont {S.}~\bibnamefont {Fairhurst}},
  \bibinfo {author} {\bibfnamefont {M.}~\bibnamefont {Hannam}}, \ and\ \bibinfo
  {author} {\bibfnamefont {P.}~\bibnamefont {Murphy}},\ }\href {\doibase
  10.1103/PhysRevD.87.024035} {\bibfield  {journal} {\bibinfo  {journal} {Phys.
  Rev. D}\ }\textbf {\bibinfo {volume} {87}},\ \bibinfo {pages} {024035}
  (\bibinfo {year} {2013})}\BibitemShut {NoStop}%
\bibitem [{\citenamefont {Sengupta}\ \emph {et~al.}(2003)\citenamefont
  {Sengupta}, \citenamefont {Dhurandhar},\ and\ \citenamefont
  {Lazzarini}}]{Sengupta:2003wk}%
  \BibitemOpen
  \bibfield  {author} {\bibinfo {author} {\bibfnamefont {A.~S.}\ \bibnamefont
  {Sengupta}}, \bibinfo {author} {\bibfnamefont {S.}~\bibnamefont
  {Dhurandhar}}, \ and\ \bibinfo {author} {\bibfnamefont {A.}~\bibnamefont
  {Lazzarini}},\ }\href {\doibase 10.1103/PhysRevD.67.082004} {\bibfield
  {journal} {\bibinfo  {journal} {Phys. Rev.}\ }\textbf {\bibinfo {volume}
  {D67}},\ \bibinfo {pages} {082004} (\bibinfo {year} {2003})},\ \Eprint
  {http://arxiv.org/abs/gr-qc/0301025} {arXiv:gr-qc/0301025 [gr-qc]}
  \BibitemShut {NoStop}%
\bibitem [{\citenamefont {Ohme}\ \emph {et~al.}(2013)\citenamefont {Ohme},
  \citenamefont {Nielsen}, \citenamefont {Keppel},\ and\ \citenamefont
  {Lundgren}}]{Ohme:2013nsa}%
  \BibitemOpen
  \bibfield  {author} {\bibinfo {author} {\bibfnamefont {F.}~\bibnamefont
  {Ohme}}, \bibinfo {author} {\bibfnamefont {A.~B.}\ \bibnamefont {Nielsen}},
  \bibinfo {author} {\bibfnamefont {D.}~\bibnamefont {Keppel}}, \ and\ \bibinfo
  {author} {\bibfnamefont {A.}~\bibnamefont {Lundgren}},\ }\href {\doibase
  10.1103/PhysRevD.88.042002} {\bibfield  {journal} {\bibinfo  {journal} {Phys.
  Rev.}\ }\textbf {\bibinfo {volume} {D88}},\ \bibinfo {pages} {042002}
  (\bibinfo {year} {2013})},\ \Eprint {http://arxiv.org/abs/1304.7017}
  {arXiv:1304.7017 [gr-qc]} \BibitemShut {NoStop}%
\bibitem [{\citenamefont {Pan}\ \emph {et~al.}(2004)\citenamefont {Pan},
  \citenamefont {Buonanno}, \citenamefont {Chen},\ and\ \citenamefont
  {Vallisneri}}]{Pan:2003qt}%
  \BibitemOpen
  \bibfield  {author} {\bibinfo {author} {\bibfnamefont {Y.}~\bibnamefont
  {Pan}}, \bibinfo {author} {\bibfnamefont {A.}~\bibnamefont {Buonanno}},
  \bibinfo {author} {\bibfnamefont {Y.-b.}\ \bibnamefont {Chen}}, \ and\
  \bibinfo {author} {\bibfnamefont {M.}~\bibnamefont {Vallisneri}},\ }\href
  {\doibase 10.1103/PhysRevD.69.104017, 10.1103/PhysRevD.74.029905} {\bibfield
  {journal} {\bibinfo  {journal} {Phys. Rev.}\ }\textbf {\bibinfo {volume}
  {D69}},\ \bibinfo {pages} {104017} (\bibinfo {year} {2004})},\ \Eprint
  {http://arxiv.org/abs/gr-qc/0310034} {arXiv:gr-qc/0310034 [gr-qc]}
  \BibitemShut {NoStop}%
\bibitem [{\citenamefont {Marsat}\ \emph {et~al.}(2013)\citenamefont {Marsat},
  \citenamefont {Bohe}, \citenamefont {Faye},\ and\ \citenamefont
  {Blanchet}}]{Marsat:2012fn}%
  \BibitemOpen
  \bibfield  {author} {\bibinfo {author} {\bibfnamefont {S.}~\bibnamefont
  {Marsat}}, \bibinfo {author} {\bibfnamefont {A.}~\bibnamefont {Bohe}},
  \bibinfo {author} {\bibfnamefont {G.}~\bibnamefont {Faye}}, \ and\ \bibinfo
  {author} {\bibfnamefont {L.}~\bibnamefont {Blanchet}},\ }\href {\doibase
  10.1088/0264-9381/30/5/055007} {\bibfield  {journal} {\bibinfo  {journal}
  {Class. Quant. Grav.}\ }\textbf {\bibinfo {volume} {30}},\ \bibinfo {pages}
  {055007} (\bibinfo {year} {2013})},\ \Eprint {http://arxiv.org/abs/1210.4143}
  {arXiv:1210.4143 [gr-qc]} \BibitemShut {NoStop}%
\bibitem [{\citenamefont {Wade}\ \emph {et~al.}(2013)\citenamefont {Wade},
  \citenamefont {Creighton}, \citenamefont {Ochsner},\ and\ \citenamefont
  {Nielsen}}]{PhysRevD.88.083002}%
  \BibitemOpen
  \bibfield  {author} {\bibinfo {author} {\bibfnamefont {M.}~\bibnamefont
  {Wade}}, \bibinfo {author} {\bibfnamefont {J.~D.~E.}\ \bibnamefont
  {Creighton}}, \bibinfo {author} {\bibfnamefont {E.}~\bibnamefont {Ochsner}},
  \ and\ \bibinfo {author} {\bibfnamefont {A.~B.}\ \bibnamefont {Nielsen}},\
  }\href {\doibase 10.1103/PhysRevD.88.083002} {\bibfield  {journal} {\bibinfo
  {journal} {Phys. Rev. D}\ }\textbf {\bibinfo {volume} {88}},\ \bibinfo
  {pages} {083002} (\bibinfo {year} {2013})}\BibitemShut {NoStop}%
\bibitem [{mon()}]{mongodb}%
  \BibitemOpen
  \href@noop {} {\enquote {\bibinfo {title} {{Mongo DB}},}\ }\bibinfo
  {howpublished} {\url{http://www.mongodb.org}}\BibitemShut {NoStop}%
\bibitem [{\citenamefont {Nielsen}(2013)}]{Nielsen:2012sb}%
  \BibitemOpen
  \bibfield  {author} {\bibinfo {author} {\bibfnamefont {A.~B.}\ \bibnamefont
  {Nielsen}},\ }\href {\doibase 10.1088/0264-9381/30/7/075023} {\bibfield
  {journal} {\bibinfo  {journal} {Class.Quant.Grav.}\ }\textbf {\bibinfo
  {volume} {30}},\ \bibinfo {pages} {075023} (\bibinfo {year} {2013})},\
  \Eprint {http://arxiv.org/abs/1203.6603} {arXiv:1203.6603 [gr-qc]}
  \BibitemShut {NoStop}%
\bibitem [{\citenamefont {Ozel}\ \emph {et~al.}(2010)\citenamefont {Ozel},
  \citenamefont {Psaltis}, \citenamefont {Narayan},\ and\ \citenamefont
  {McClintock}}]{Ozel:2010su}%
  \BibitemOpen
  \bibfield  {author} {\bibinfo {author} {\bibfnamefont {F.}~\bibnamefont
  {Ozel}}, \bibinfo {author} {\bibfnamefont {D.}~\bibnamefont {Psaltis}},
  \bibinfo {author} {\bibfnamefont {R.}~\bibnamefont {Narayan}}, \ and\
  \bibinfo {author} {\bibfnamefont {J.~E.}\ \bibnamefont {McClintock}},\ }\href
  {\doibase 10.1088/0004-637X/725/2/1918} {\bibfield  {journal} {\bibinfo
  {journal} {Astrophys.J.}\ }\textbf {\bibinfo {volume} {725}},\ \bibinfo
  {pages} {1918} (\bibinfo {year} {2010})},\ \Eprint
  {http://arxiv.org/abs/1006.2834} {arXiv:1006.2834 [astro-ph.GA]} \BibitemShut
  {NoStop}%
\bibitem [{\citenamefont {Kiziltan}\ \emph {et~al.}(2013)\citenamefont
  {Kiziltan}, \citenamefont {Kottas}, \citenamefont {De~Yoreo},\ and\
  \citenamefont {Thorsett}}]{Kiziltan:2013oja}%
  \BibitemOpen
  \bibfield  {author} {\bibinfo {author} {\bibfnamefont {B.}~\bibnamefont
  {Kiziltan}}, \bibinfo {author} {\bibfnamefont {A.}~\bibnamefont {Kottas}},
  \bibinfo {author} {\bibfnamefont {M.}~\bibnamefont {De~Yoreo}}, \ and\
  \bibinfo {author} {\bibfnamefont {S.~E.}\ \bibnamefont {Thorsett}},\ }\href
  {\doibase 10.1088/0004-637X/778/1/66} {\bibfield  {journal} {\bibinfo
  {journal} {Astrophys.J.}\ }\textbf {\bibinfo {volume} {778}},\ \bibinfo
  {pages} {66} (\bibinfo {year} {2013})},\ \Eprint
  {http://arxiv.org/abs/1309.6635} {arXiv:1309.6635 [astro-ph.SR]} \BibitemShut
  {NoStop}%
\bibitem [{\citenamefont {Finn}\ and\ \citenamefont
  {Chernoff}(1993)}]{Finn:1992xs}%
  \BibitemOpen
  \bibfield  {author} {\bibinfo {author} {\bibfnamefont {L.~S.}\ \bibnamefont
  {Finn}}\ and\ \bibinfo {author} {\bibfnamefont {D.~F.}\ \bibnamefont
  {Chernoff}},\ }\href {\doibase 10.1103/PhysRevD.47.2198} {\bibfield
  {journal} {\bibinfo  {journal} {Phys. Rev.}\ }\textbf {\bibinfo {volume}
  {D47}},\ \bibinfo {pages} {2198} (\bibinfo {year} {1993})},\ \Eprint
  {http://arxiv.org/abs/gr-qc/9301003} {arXiv:gr-qc/9301003 [gr-qc]}
  \BibitemShut {NoStop}%
\bibitem [{\citenamefont {Blackburn}\ \emph {et~al.}(2008)\citenamefont
  {Blackburn}, \citenamefont {Cadonati}, \citenamefont {Caride}, \citenamefont
  {Caudill}, \citenamefont {Chatterji} \emph {et~al.}}]{Blackburn:2008ah}%
  \BibitemOpen
  \bibfield  {author} {\bibinfo {author} {\bibfnamefont {L.}~\bibnamefont
  {Blackburn}}, \bibinfo {author} {\bibfnamefont {L.}~\bibnamefont {Cadonati}},
  \bibinfo {author} {\bibfnamefont {S.}~\bibnamefont {Caride}}, \bibinfo
  {author} {\bibfnamefont {S.}~\bibnamefont {Caudill}}, \bibinfo {author}
  {\bibfnamefont {S.}~\bibnamefont {Chatterji}},  \emph {et~al.},\ }\href
  {\doibase 10.1088/0264-9381/25/18/184004} {\bibfield  {journal} {\bibinfo
  {journal} {Class.Quant.Grav.}\ }\textbf {\bibinfo {volume} {25}},\ \bibinfo
  {pages} {184004} (\bibinfo {year} {2008})},\ \Eprint
  {http://arxiv.org/abs/0804.0800} {arXiv:0804.0800 [gr-qc]} \BibitemShut
  {NoStop}%
\bibitem [{\citenamefont {Slutsky}\ \emph {et~al.}(2010)\citenamefont
  {Slutsky}, \citenamefont {Blackburn}, \citenamefont {Brown}, \citenamefont
  {Cadonati}, \citenamefont {Cain} \emph {et~al.}}]{Slutsky:2010ff}%
  \BibitemOpen
  \bibfield  {author} {\bibinfo {author} {\bibfnamefont {J.}~\bibnamefont
  {Slutsky}}, \bibinfo {author} {\bibfnamefont {L.}~\bibnamefont {Blackburn}},
  \bibinfo {author} {\bibfnamefont {D.}~\bibnamefont {Brown}}, \bibinfo
  {author} {\bibfnamefont {L.}~\bibnamefont {Cadonati}}, \bibinfo {author}
  {\bibfnamefont {J.}~\bibnamefont {Cain}},  \emph {et~al.},\ }\href {\doibase
  10.1088/0264-9381/27/16/165023} {\bibfield  {journal} {\bibinfo  {journal}
  {Class.Quant.Grav.}\ }\textbf {\bibinfo {volume} {27}},\ \bibinfo {pages}
  {165023} (\bibinfo {year} {2010})},\ \Eprint {http://arxiv.org/abs/1004.0998}
  {arXiv:1004.0998 [gr-qc]} \BibitemShut {NoStop}%
\bibitem [{\citenamefont {Dent}\ and\ \citenamefont
  {Veitch}(2014)}]{PhysRevD.89.062002}%
  \BibitemOpen
  \bibfield  {author} {\bibinfo {author} {\bibfnamefont {T.}~\bibnamefont
  {Dent}}\ and\ \bibinfo {author} {\bibfnamefont {J.}~\bibnamefont {Veitch}},\
  }\href {\doibase 10.1103/PhysRevD.89.062002} {\bibfield  {journal} {\bibinfo
  {journal} {Phys. Rev. D}\ }\textbf {\bibinfo {volume} {89}},\ \bibinfo
  {pages} {062002} (\bibinfo {year} {2014})}\BibitemShut {NoStop}%
\end{thebibliography}%
\bibliographystyle{apsrev4-1}

\end{document}